\documentclass[12pt]{iopart} 
 
\usepackage{iopams}  

\usepackage{bm}
\usepackage[dvips]{graphicx}

\newcommand{\bmJ}{{\bm J}}
\newcommand{\bmv}{{\bm v}}

\newcommand{\ui}{{\rm i}} 
\newcommand{\bmr}{{\bm r}}
\newcommand{\bmR}{{\bm R}}
\newcommand{\bms}{{\bm s}}
\newcommand{\bmS}{{\bm S}}
\newcommand{\bmq}{{\bm q}}
\newcommand{\bmQ}{{\bm Q}}
\newcommand{\bmk}{{\bm k}}
\newcommand{\bmK}{{\bm K}}
\newcommand{\bmp}{{\bm p}}
\newcommand{\kB}{k_{\rm B}}

\begin{document}

\title[Theory of the Spin Seebeck Effect]{Theory of the Spin Seebeck Effect}

\author{Hiroto Adachi$^{1,2}$, Ken-ichi Uchida$^{3,4}$, Eiji Saitoh$^{3,5,1,2}$, and Sadamichi Maekawa$^{1,2}$} 

\address{$^1$ Advanced Science Research Center, Japan Atomic Energy Agency, Tokai 319-1195, Ibaraki, Japan} 
\address{$^2$ CREST, Japan Science and Technology Agency, Sanbancho, Tokyo 102-0075, Japan}
\address{$^3$ Institute for Materials Research, Tohoku University, Sendai 980-8577, Japan}
\address{$^4$ PRESTO, Japan Science and Technology Agency, Kawaguchi, Saitama 332-0012, Japan} 
\address{$^5$ WPI Advanced Institute for Materials Research, Tohoku University, Sendai 980-8577, Japan} 

\ead{adachi.hiroto@jaea.go.jp}

\begin{abstract}
The spin Seebeck effect refers to the generation of a spin voltage caused by a temperature gradient in a ferromagnet, which enables the thermal injection of spin currents from the ferromagnet into an attached nonmagnetic metal over a macroscopic scale of several millimeters. The inverse spin Hall effect converts the injected spin current into a transverse charge voltage, thereby producing electromotive force as in the conventional charge Seebeck device. Recent theoretical and experimental efforts have shown that the magnon and phonon degrees of freedom play crucial roles in the spin Seebeck effect. In this article, we present the theoretical basis for understanding the spin Seebeck effect and briefly discuss other thermal spin effects. 
\end{abstract}

\maketitle

\section{Introduction}

Generation of electromotive force by a temperature gradient has been known for many years as the Seebeck effect~\cite{Goldsmid-text}. In recent years, a spin analogue of the Seebeck effect has drawn much attention in the field of spintronics, because replacing charge transport with spin transport in modern solid state devices is a major issue in the spintronics community. More than two decades ago, Johnson and Silsbee~\cite{Johnson87} published a seminal theoretical study, in which they generalized the interfacial thermoelectric effect to include spin transport phenomena. Because their framework implicitly relies on a spin transport carried by spin-polarized conduction electrons, the phenomenon discussed in Ref.~\cite{Johnson87} should be classified as a ``spin-dependent'' Seebeck effect from this perspective. The field of thermal spintronics is sometimes called spin caloritronics~\cite{Bauer12}. An experiment reported in 2008 put a new twist on spin caloritronics, because understanding of that experiment requires a framework other than the ``spin-dependent'' Seebeck effect. 

In 2008, Uchida {\it et al.} demonstrated that when a ferromagnetic film is placed under the influence of a temperature gradient, a spin current is injected from the ferromagnetic film into attached nonmagnetic metals with the signal observed over a macroscopic scale of several millimeters~\cite{Uchida08}. This phenomenon, termed the spin Seebeck effect, surprised the community because the length scale seen in the experiment was extraordinarily longer than the spin-flip diffusion length of conduction electrons, suggesting that the conduction electrons in the ferromagnet are irrelevant to the phenomenon. Subsequently, the spin Seebeck effect was observed in various materials ranging from the metallic ferromagnets Co$_2$MnSi~\cite{Bosu11} to the semiconducting ferromagnet (Ga,Mn)As~\cite{Jaworski10}, and even in the insulating magnets LaY$_2$Fe$_5$O$_{12}$~\cite{Uchida10a} and (Mn,Zn)Fe$_2$O$_4$~\cite{Uchida10c}. These observations have established the spin Seebeck effect as a universal aspect of ferromagnets. 

In a spin Seebeck device, the spin current injected into an attached nonmagnetic metal is converted into a transverse charge voltage with the help of the inverse spin Hall effect~\cite{Saitoh06,Valenzuela06,Kimura07}. Therefore, the spin Seebeck effect enables the generation of electromotive force from the temperature gradient as in conventional charge Seebeck devices. What is new in the spin Seebeck device is that it has a scalability different from that of conventional charge Seebeck devices, in that the output power is proportional to the length perpendicular to the temperature gradient. In addition, the paths of the heat current and charge current are separated in the spin Seebeck device in contrast to the charge Seeebck device, such that the spin Seebeck device could be a new way to enhance the thermoelectric efficiency. Because of these new features, an attempt is already underway to develop a new spin Seebeck thermoelectric device~\cite{Kirihara12,Uchida11c,Uchida12b}. 

As is inferred from the fact that the spin Seebeck effect occurs even in an insulating magnet~\cite{Uchida10a}, this phenomenon cannot be described by the the ``spin-dependent'' Seebeck framework proposed by Johnson and Silsbee~\cite{Johnson87}. Instead, we need several new ideas and notions. In this article, we introduce basic ideas to understand the spin Seebeck effect. In addition, we present a brief summary of other thermo-spin phenomena.

\section{Spin Current}

The spin Seebeck effect is a long-range thermal injection of the spin current from a ferromagnet into an attached nonmagnetic metal. Therefore, knowledge on the spin current is indispensable for understanding the spin Seebeck effect. In spin-orbit coupled systems, the spin is a non-conserved quantity, and hence there have been a number of discussions on the proper definition of spin currents in such systems~\cite{Shi06,Nikolic06}. We do not discuss this subtle problem in this article, but here we present a simple argument. Let us consider the following definition of a spin current ${\bm J}_{\rm s}$: 
\begin{equation}
  \bmJ_{\rm s} = \sum_\bmk s^z_\bmk \bmv_\bmk , 
  \label{Eq:spincurrent01}
\end{equation}
where 
$s^z_\bmk$ is the $z$-component of the spin density $\bms_\bmk$ with the $z$ axis chosen as a spin-quantizing axis, and $\bmv_\bmk$ is the velocity of elementary excitations concomitant to the spin density $\bms_\bmk$. We consider here a spin-independent velocity $\bmv_\bmk$ because we focus on a pure spin current that is unaccompanied by a charge current. 

From Eq.~(\ref{Eq:spincurrent01}) we can derive two kinds of pure spin current. The first is the so-called conduction-electron pure spin current. In this case, the $z$-component of the spin density is given by 
$s^z_\bmk= c^\dag_{\bmk,\uparrow} c_{\bmk,\uparrow} - c^\dag_{\bmk,\downarrow} c_{\bmk,\downarrow}$, 
where $c^\dag_{\bmk,\sigma}$ is the creation operator for conduction electrons with spin projection $\sigma=\uparrow,\downarrow$ and momentum $\bmk$. After taking the statistical average, the expectation value of the conduction-electron pure spin current $\bmJ^{\rm c \mathchar`- el}_{\rm s}$ is calculated to be 
\begin{equation}
  \bmJ^{\rm c \mathchar`- el}_{\rm s} = 
  \sum_\bmk {\bm v}_\bmk 
  \Big( \langle c^\dag_{\bmk,\uparrow} c_{\bmk,\uparrow} \rangle 
  - \langle c^\dag_{\bmk,\downarrow} c_{\bmk,\downarrow} \rangle \Big), 
\end{equation}
where ${\bm v}_\bmk$ is the velocity of conduction electrons. From this expression, we see that an asymmetry between the up-spin population and the down-spin population is necessary to obtain a nonzero conduction-electron pure spin current. 

The second type of pure spin current is the so-called magnon spin current. In this case the $z$-component of the spin density is given by $s^z_\bmk= S_0- b^\dag_\bmk b_\bmk$, where $b^\dag_\bmk$ is the creation operator for magnons with momentum $\bmk$. Substituting this into Eq.~(\ref{Eq:spincurrent01}) and taking the statistical average, the expectation value of the magnon pure spin current $\bmJ^{\rm mag}_{\rm s}$ is given by 
\begin{equation}
  \bmJ^{\rm mag}_{\rm s} 
  = 
  -\frac{1}{2} \sum_{\bmk} \bmv_{\bmk}
  \Big( \langle b^\dag_{\bmk} b_{\bmk} \rangle 
  - \langle b^\dag_{-\bmk} b_{-\bmk} \rangle \Big), 
\end{equation}
where $\bmv_\bmk$ is the magnon velocity, and we have used the relation $\bmv_{-\bmk}= - \bmv_\bmk$. From this expression, we see that an asymmetry between the left-moving population and the right-moving population is necessary to obtain a nonzero magnon spin current. 

\begin{figure}[h]
\begin{center}\scalebox{0.65}[0.65]{\includegraphics{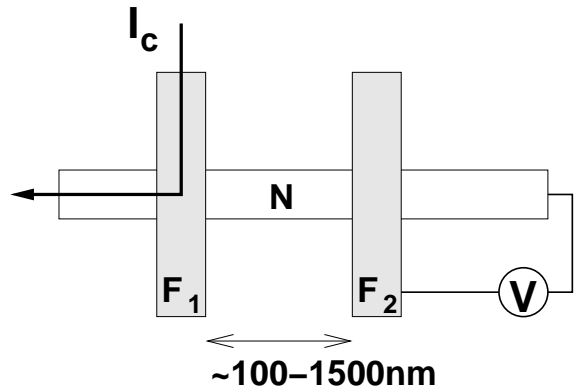}}\end{center} 
\caption{Schematic of a device that injects and detects the conduction-electron spin current~\cite{Jedema01}. } 
\label{fig_Nonlocal01}
\end{figure}

\begin{figure}[h]
\begin{center}\scalebox{0.35}[0.35]{\includegraphics{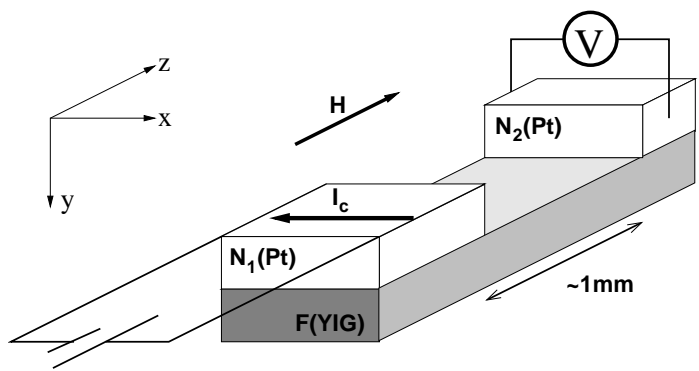}}\end{center} 
\caption{Schematic of a device that injects and detects the magnon spin current~\cite{Kajiwara10}.} 
\label{fig_Kajiwara01}
\end{figure}

These two spin currents can be detected experimentally in the following way. For the conduction-electron spin current $\bmJ^{\rm c \mathchar`- el}_{\rm s}$, the method of nonlocal spin injection and detection is used~\cite{Jedema01,Jedema02}. In the device shown in Fig.~\ref{fig_Nonlocal01}, a charge current $I_c$ is applied across the interface between a metallic ferromagnet $F_1$ and a nonmagnetic metal $N$. Because the conduction electrons in $F_1$ are spin polarized, a spin-polarized current is injected from $F_1$ into $N$, which creates a spin accumulation at the interface between $F_1$ and $N$. Then, because there is no charge current flowing to the right side of $F_1$, the spin accumulation at the $F_1/N$ interface diffuses to the right in the form of a conduction-electron spin current. The signal of the conduction-electron spin current is detected through the second metallic ferromagnet $F_2$ by measuring the electric voltage as shown in Fig.~\ref{fig_Nonlocal01}. If and only if there is a spin accumulation at the $F_2/N$ interface, the electrochemical potential at the $F_2/N$ interface is influenced by whether or not the magnetization in $F_2$ is parallel to that in $F_1$ (for more details, see Ref.~\cite{Takahashi08}). 

For a magnon spin current $\bmJ^{\rm mag}_{\rm s}$, an insulating magnet is used to eliminate the contribution from the conduction-electron spin current~\cite{Kajiwara10}. In Fig.~\ref{fig_Kajiwara01}, two platinum films are put on top of a yttrium iron garnet (YIG) film. The first Pt film ($N_1$) acts as a spin current injector with the help of the spin Hall effect (see the next section). The spin current injected from $N_1$ exerts spin torque on the localized magnetic moment at the $N_1/F$ interface. Owing to the spin torque, the magnetization at the $N_1/F$ interface starts to precess and induces a spin current. Then the spin current propagates through $F$ in the form of a magnon spin current $\bmJ^{\rm mag}_{\rm s}$. When the magnon spin current propagates from the $N_1/F$ interface to the $N_2/F$ interface and the localized spins at the interface are excited, the spin current is injected from $F$ into $N_2$ owing to the $s$-$d$ exchange interaction at the interface~\cite{Takahashi09}. The spin current thus injected can be detected electrically via the inverse spin Hall effect (see the next section). 

The important point here is the difference in the decay lengths between the conduction-electron spin current and the magnon spin current. The conduction-electron spin current $\bmJ^{\rm c \mathchar`-el}_{\rm s}$ decays over $100$-$1500$ nm in metals depending on the strength of the spin-orbit interaction~\cite{Jedema01,Jedema02,Takahashi08}. On the other hand, the magnon spin current can sometimes propagate over a macroscopic length scale of a millimeter, which was indeed observed in Ref.~\cite{Kajiwara10}.

\section{Spin Hall Effect}
\label{Sec:SHE}

The spin Hall effect refers to the appearance of a nonzero spin current in the direction transverse to the applied charge current. In the spin Seebeck effect, the spin current injected from a ferromagnet into an attached nonmagnetic metal is converted into a transverse charge voltage via the reverse of the spin Hall effect, the so-called inverse spin Hall effect~\cite{Saitoh06,Valenzuela06,Kimura07}. Namely, the inverse spin Hall effect is used for electrical detection of the spin Seebeck effect. There are already a number of publications on the spin Hall effect in the literature, and we recommend Ref.~\cite{Takahashi08} for readers interested in the detailed derivation of the spin Hall effect. 

The basic idea of the spin Hall effect~\cite{Dyakonov71,Hirsch99} is as follows. It is known that, in the presence of the spin-orbit interaction, a scattered electron acquires a spin polarization with the polarization vector $\widehat{\bm \sigma}$ given by 
\begin{equation}
  \widehat{\bm \sigma} \propto \widehat{\bmk}_{\rm in} \times \widehat{\bmk}_{\rm out}, 
  \label{Eq:SHE01}
\end{equation}
where $\widehat{\bmk}_{\rm in}$ and $\widehat{\bmk}_{\rm out}$ are the incident and scattered wave vectors. By multiplying 
both sides of Eq.~(\ref{Eq:SHE01}) by the vector $\widehat{\bmk}_{\rm in}$, we see that the component of the scattered wave vector perpendicular to the incident wave vector, 
i.e.,  
${\bmk}^\perp_{\rm out} =  
\widehat{\bmk}_{\rm out}- (\widehat{\bmk}_{\rm in} \cdot \widehat{\bmk}_{\rm out}) 
\widehat{\bmk}_{\rm in}$,  
is given by 
\begin{equation}
  {\bmk}^\perp_{\rm out} \propto \widehat{\bm \sigma} \times \widehat{\bmk}_{\rm in}. 
  \label{Eq:k_perp01}
\end{equation}
This equation means that the scattered vector is determined by the spin state and wave vector of the incident electrons. Macroscopically the spin Hall effect can be expressed as~\cite{Takahashi08,Zhang01,Engel05,Tse06} 
\begin{equation}
  \widetilde{\bmJ}_{\rm s} = \theta_H \widehat{\bm \sigma} \times \bmJ_{\rm c}, 
  \label{Eq:SHE02}
\end{equation}
while the inverse spin Hall effect is expressed as 
\begin{equation}
  \bmJ_{\rm c} = \theta_H \widehat{\bm \sigma} \times \widetilde{\bmJ}_{\rm s}, 
  \label{Eq:ISHE01}
\end{equation}
where $\theta_H$ is the spin Hall angle, $\widehat{\bm \sigma}$ denotes the direction of the spin polarization, and $\widetilde{\bmJ}_{\rm s}=e \bmJ_{\rm s}$ with $e$ being the electronic charge. 

\begin{figure}[t]
\begin{center}\scalebox{0.18}[0.18]{\includegraphics{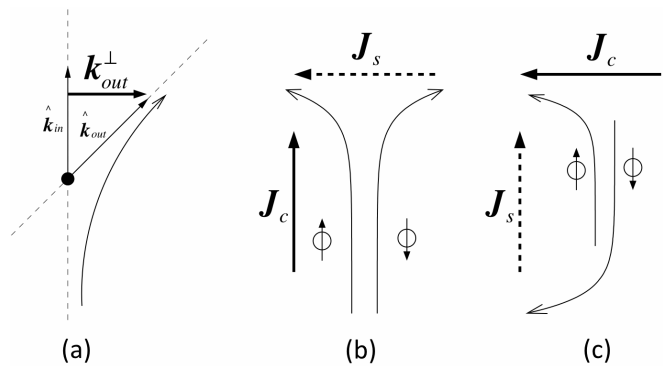}}\end{center} 
\caption{(a) Definition of 
$\bmk^\perp_{\rm out}= 
\widehat{\bmk}_{\rm out}- (\widehat{\bmk}_{\rm in} \cdot \widehat{\bmk}_{\rm out}) 
\widehat{\bmk}_{\rm in}$ 
appearing in Eq.~(\ref{Eq:k_perp01}), where $\widehat{\bmk}_{\rm in}$ and $\widehat{\bmk}_{\rm out}$ are the incident and scattered wave vectors. (b) Schematic of the spin Hall effect. The charge current $\bmJ_{\rm c}$ is converted into the transverse spin current $\bmJ_{\rm s}$. (c) Schematic illustration of the inverse spin Hall effect. The spin current $\bmJ_{\rm s}$ is converted into the transverse charge current $\bmJ_{\rm c}$. The spin-quantizing axis is perpendicular to the plane of the sheet. }
\label{fig_SHE-cartoon01}
\end{figure}

We now explain how the spin Hall effect works by taking the experiment of injection and detection of a magnon spin current via the spin Hall effect~\cite{Kajiwara10} (Fig.~\ref{fig_Kajiwara01}) as an example. Here, the nonmagnetic metal $N_1$ is used as a spin-current injector by means of the spin Hall effect. In $N_1$, a charge current $\bmJ_c$ is applied parallel to the $x$ direction. Then the spin current $\bmJ_{\rm s}$ ($\parallel \widehat{\bm y}$) across the $N_1/F$ interface that is generated by the spin Hall effect has a spin polarization along the $z$ axis owing to Eq.~(\ref{Eq:SHE02}). This spin current $\bmJ_{\rm s}$ creates a spin accumulation ${\bm \mu}$ ($\parallel \widehat{\bm z}$) at the $N_1/F$ interface, and through the $s$-$d$ exchange interaction at the interface~\cite{Zhang04} it exerts a spin torque on the magnetization ${\bm M}$ at the $N_1/F$ interface in the form ${\bm T} \propto {\bm M} \times ({\bm M} \times {\bm \mu})$~\cite{Slonczewski96,Berger96,Zhang04}. This torque excites a magnon spin current in the ferromagnet. 

The nonmagnetic metal $N_2$ is used to detect the magnon spin current by means of the inverse spin Hall effect. When the magnon spin current propagates from the $N_1/F$ interface to the $N_2/F$ interface, it injects spins from $F$ into $N_2$ with a spin polarization along the $z$ axis, again owing to the $s$-$d$ exchange interaction at the interface~\cite{Takahashi08}. The injected spins polarized parallel to the $z$ axis diffuse along the $y$ axis, and are converted into a charge current along the $x$ axis owing to the inverse spin Hall effect Eq.~(\ref{Eq:ISHE01}). Therefore, the magnon spin current is detected as a charge voltage as shown in Fig.~\ref{fig_Kajiwara01}. The inverse spin Hall effect plays an important role in electrically detecting the spin Seebeck effect.

\section{Spin Seebeck Effect}

The spin Seebeck effect is the generation of a spin voltage caused by a temperature gradient in a ferromagnet. Here, the spin voltage is a potential for electrons' spin to drive spin currents. More concretely, when a nonmagnetic metal is attached on top of a material with a finite spin voltage, a nonzero spin injection is obtained. In this section, we first present a brief summary of the spin Seebeck effect, and then show the experimental details of this effect.

\subsection{Brief summary of the spin Seebeck effect} 
\begin{figure}[t]
\begin{center}\scalebox{0.4}[0.4]{\includegraphics{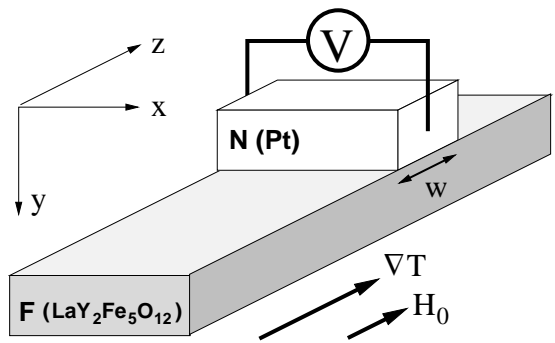}}\end{center} 
\caption{Schematic of the experimental setup for observing the spin Seebeck effect~\cite{Uchida08}.} 
\label{fig_SSEsetup01} 
\end{figure}

Figure~\ref{fig_SSEsetup01} shows the experimental setup for observing the spin Seebeck effect in a magnetic insulator LaY$_2$Fe$_5$O$_{12}$~\cite{Uchida10a}. Here a Pt strip is attached on top of a LaY$_2$Fe$_5$O$_{12}$ film in a static magnetic field ${\bm H}_0 = H_0 {\bm \hat{\bm z}}$ ($\gg$ anisotropy field), which aligns the localized magnetic moment along ${\bm \hat{\bm z}}$. A temperature gradient ${\bm \nabla}T $ is applied along the $z$-axis, which induces a spin voltage across the LaY$_2$Fe$_5$O$_{12}$/Pt interface. Then this spin voltage injects a spin current $I_s$ into the Pt strip (or ejects it from the Pt strip). A part of the injected/ejected spin current $I_s$ is converted into a charge voltage through the inverse spin Hall effect~\cite{Saitoh06,Valenzuela06,Kimura07}: 
\begin{equation}
V = \theta_H (|e|I_s)(\rho/w), 
\label{Eq:ISHE02} 
\end{equation}
where $|e|$, $\theta_H$, $\rho$, and $w$ are the absolute value of electron charge, the spin Hall angle, the electrical resistivity, and the width of the Pt strip (see Fig.~\ref{fig_SSEsetup01}). Hence, the observed charge voltage $V$ is a measure of the injected/ejected spin current $I_s$. By using this configuration, the spin Seebeck effect is observed not only in ferromagnetic metals (NiFe alloys~\cite{Uchida08} and Co$_2$MnSi~\cite{Bosu11}), but also in ferromagnetic semiconductors ((Ga,Mn)As)~\cite{Jaworski10} and insulators (LaY$_2$Fe$_5$O$_{12}$~\cite{Uchida10a} and (Mn,Zn)Fe$_2$O$_4$~\cite{Uchida10c}). 

\begin{figure}[t]
\begin{center}
\includegraphics[width=8cm]{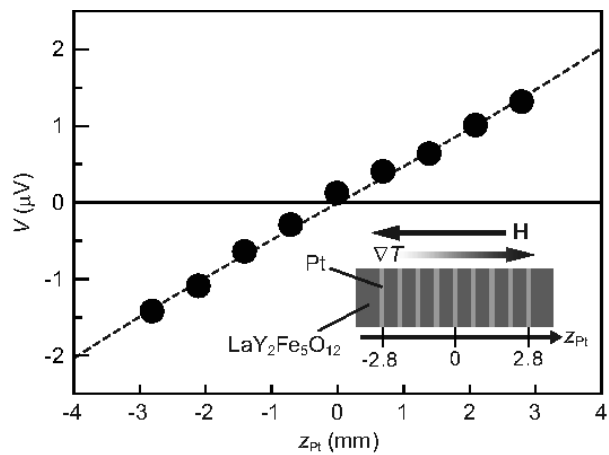}
\caption{
Dependence of the observed voltage $V$ on $z_{\rm Pt}$, the displacement of the Pt wire from the center of the LaY$_2$Fe$_5$O$_{12}$ layer along the $z$ direction, in the LaY$_2$Fe$_5$O$_{12}$/Pt sample at $\Delta T = 20~\textrm{K}$. }
\label{fig:transverseSSE2}
\end{center}
\end{figure}

As shown in Fig.~\ref{fig:transverseSSE2}, the spatial dependence of the spin Seebeck effect can be measured by changing the position of the Pt strip. Note that the signal has a quasi-linear spatial dependence, with the signal changing signs at both ends of the sample and vanishing at the center of the sample. 

It has been shown that the conduction electrons alone cannot explain the spin Seebeck effect, because the conduction electrons' short spin-flip diffusion length ($\sim $ several nanometers in a NiFe alloy) fails to explain the long length scale ($\sim$ several millimeters) observed in experiments~\cite{Hatami10,Nunner11}. This interpretation is further supported by the following two experiments. As we have already discussed, it was demonstrated in Ref.~\cite{Kajiwara10} using a ferromagnetic insulator YIG that spin currents can be carried by magnon excitations. Subsequently, it was reported that, despite the absence of conduction electrons, the spin Seebeck effect can be observed in LaY$_2$Fe$_5$O$_{12}$, a magnetic insulator~\cite{Uchida10a}. These experiments suggest that, contrary to the conventional view of the last two decades that the spin current is carried by conduction electrons~\cite{Maekawa06}, the magnon can be a carrier for the spin Seebeck effect. 

Now there is a consensus that the spin Seebeck effect is caused by a nonequilibrium between the magnon system in the ferromagnet and the conduction electron system in the nonmagnetic metal. In certain situations, both the nonequilibrium magnons and the nonequilibrium phonons play an important role. 

Finally we note that although there is a possibility that the spin Seebeck effect in a Pt/insulating magnet hybrid system might be contaminated by the anomalous Nernst effect because of a strong magnetic proximity effect of Pt at the Pt/insulating magnet interface~\cite{Huang12}, recent experimental demonstration confirms that such a contribution is negligibly small in a Pt/YIG system~\cite{Kikkawa12}.

\subsection{Experimental details of the spin Seebeck effect}

Here we show experimental data on the spin Seebeck effect in a LaY$_2$Fe$_5$O$_{12}$/Pt sample. The sample consists of a LaY$_2$Fe$_5$O$_{12}$ film with Pt wires attached to the top surface. A single-crystal LaY$_2$Fe$_5$O$_{12}$ (111) film with a thickness of 3.9 $\mu$m was grown on a Gd$_3$Ga$_5$O$_{12}$ (111) substrate by liquid phase epitaxy, where the surface of the LaY$_2$Fe$_5$O$_{12}$ layer had an 8$\times$4 mm$^2$ rectangular shape. Two (or more) 15-nm-thick Pt wires were then sputtered in an Ar atmosphere on the top of the LaY$_2$Fe$_5$O$_{12}$ film. The length and width of the Pt wires were 4 mm and 0.1 mm, respectively. 

\begin{figure}[t]
\begin{center}\includegraphics[width=14cm]{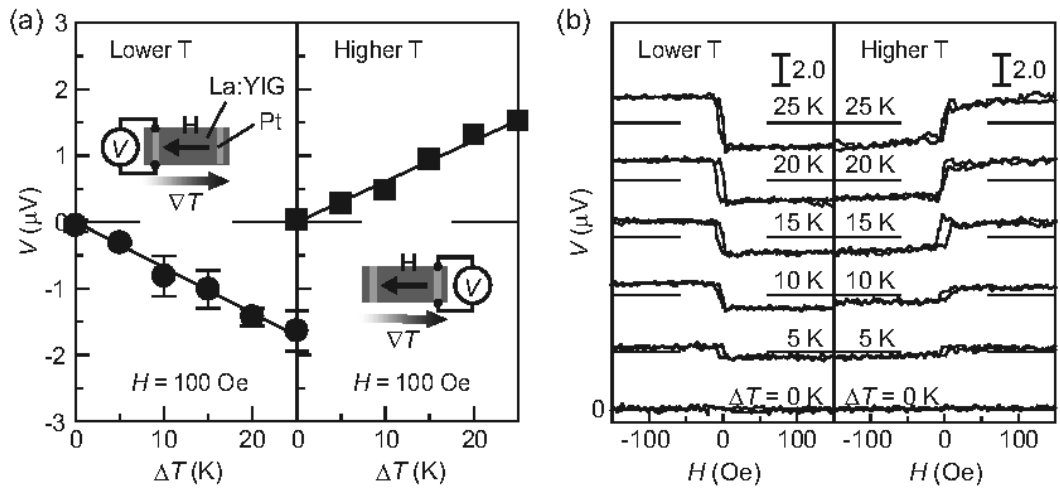}\end{center}
\caption{(a) $\Delta T$ dependence of $V$ in the LaY$_2$Fe$_5$O$_{12}$/Pt sample at $H=100~\textrm{Oe}$, measured when the Pt wires were attached near the lower-temperature (300 K) and higher-temperature (300 K+$\Delta T$) ends of the LaY$_2$Fe$_5$O$_{12}$ layer. (b) $H$ dependence of $V$ in the LaY$_2$Fe$_5$O$_{12}$/Pt sample for various values of $\Delta T$. }
\label{fig:transverseSSE1}
\end{figure}

Figures \ref{fig:transverseSSE1}(a) shows the voltage $V$ between the ends of the Pt wires placed near the lower- and higher-temperature ends of the LaY$_2$Fe$_5$O$_{12}$ layer as a function of the temperature difference $\Delta T$, measured when a magnetic field of $H = 100\,\textrm{Oe}$ was applied along the $z$ direction. The magnitude of $V$ is proportional to $\Delta T$ in both Pt wires. Notably, the sign of $V$ for finite values of $\Delta T$ is clearly reversed between the lower- and higher-temperature ends of the sample. This sign reversal of $V$ is characteristic behavior of the inverse spin Hall voltage induced by the spin Seebeck effect. 

As shown in Fig. \ref{fig:transverseSSE1}(b), the sign of $V$ at each end of the sample is reversed by reversing $H$. It was also verified that the $V$ signal vanishes when ${\bf H}$ is applied along the $x$ direction, which is consistent with Eq. (\ref{fig_SSEsetup01}). This $V$ signal disappears when the Pt wires are replaced by Cu wires with weak spin-orbit interaction. These results confirm that the $V$ signal observed here is due to the spin Seebeck effect in the LaY$_2$Fe$_5$O$_{12}$/Pt samples. 

\section{Linear-Response Theory of the Spin Seebeck effect} 

\subsection{Local picture of thermal spin injection by magnons}\label{Sec:Loc-Pic-Spump}

\begin{figure}[t]
\begin{center}\scalebox{0.5}[0.5]{\includegraphics{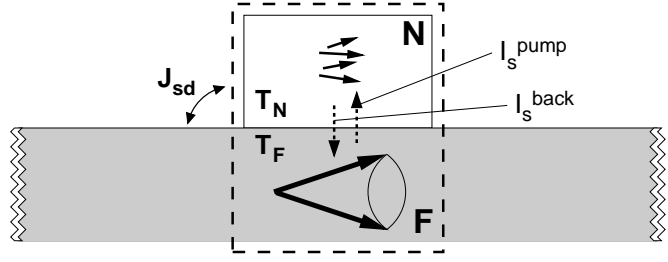}}\end{center} 
\caption{
Side-view schematic of the system considered in Sec.~\ref{Sec:Loc-Pic-Spump} for thermal spin injection. 
Here, a ferromagnet ($F$) and nonmagnetic metal ($N$) are interacting weakly through interface $s$-$d$ exchange coupling $J_{\rm sd}$, which results in the thermal injection of spin current $I_s=I_s^{\rm pump}-I_s^{\rm back}$. 
} 
\label{fig_SpinPump02}
\end{figure}

As we have already discussed, the conduction electrons in the ferromagnet are considered to be irrelevant to the spin Seebeck effect. The fact that the spin Seebeck effect is observed even in a magnetic insulator suggests that the dynamics of localized spins in the ferromagnet, or magnon, is important to the spin Seebeck effect. To understand the spin Seebeck effect from this viewpoint, we first consider a model for the thermal spin injection by localized spins (see Fig.~\ref{fig_SpinPump02}). In this model we focus on a small region encircled by the dashed line, in which a ferromagnet ($F$) with a local temperature $T_F$ and a nonmagnetic metal ($N$) with a local temperature $T_N$ are interacting weakly through interface $s$-$d$ exchange coupling $J_{\rm sd}$. For simplicity we assume that the region in question (encircled by the dashed line) is sufficiently small such that the spatial variations of any physical quantities can be neglected, and that the size of the localized spin is unity. It is also assumed that each segment is initially in local thermal equilibrium; then, the $s$-$d$ exchange interactions are switched on, and the nonequilibrium dynamics of the system is calculated. 

The physics of the ferromagnet $F$ is described by the localized moment ${\bm M}$, for which the dynamics is modeled by the Landau-Lifshitz-Gilbert equation: 
\begin{eqnarray}
  \partial_t {\bm M} &=& 
       \left[
  \gamma ( {\bm H}_{0}+ {\bm h} ) 
  -\frac{{J}_{\rm sd}}{\hbar} {\bm s}
       \right] \times 
  {\bm M}
  + \frac{\alpha}{M_s} {\bm M} \times \partial_t {\bm M}, 
  \label{Eq:LLG01} 
\end{eqnarray}
where ${\bm H}_{0}= H_0 \widehat{\bm z}$ is the external field, $\gamma$ is the gyromagnetic ratio, $\alpha$ is the Gilbert damping constant, and $M_s$ is the saturation magnetization. In the above equation, the noise field ${\bm h}$ represents the thermal fluctuations in $F$. By the fluctuation-dissipation theorem~\cite{Kubo-text,Landau-text}, ${\bm h}$ is assumed to obey the following Gaussian ensemble~\cite{Brown63}: 
\begin{eqnarray}
< h^\mu(t) > &=& 0 
\label{Eq:noise-h}
\end{eqnarray} 
and 
\begin{eqnarray}
< h^\mu(t) h^\nu (t')> &=& 
\frac{2 k_B T_F \alpha}{\gamma a_S^3 M_s} 
\delta_{\mu,\nu} \delta(t-t'), 
\label{Eq:noise-hh}
\end{eqnarray}
where $a_S^3=\hbar \gamma /M_s$ is the cell volume of the ferromagnet. 

The physics of the nonmagnetic metal $N$ is described by the itinerant spin density ${\bm s}$, and its dynamics is modeled by the Bloch equation: 
\begin{eqnarray}
   \partial_t {\bm s}&=& - \frac{1}{\tau_{\rm sf}} 
  \left( {\bm s}-s_0 \frac{{\bm M}}{M_s} \right) 
  -\frac{ {J}_{\rm sd}}{\hbar} \frac{\bm M}{M_s} \times 
  {\bm s} + {\bm l}, 
  \label{Eq:Bloch01} 
\end{eqnarray}
where $\tau_{\rm sf}$ is the spin-flip relaxation time, and $s_0= \chi_N J_{\rm sd}$ is the local equilibrium spin density~\cite{Zhang04} with the paramagnetic susceptibility $\chi_N$ in $N$. In this equation, the noise source ${\bm l}$ is introduced~\cite{Ma75} as a Gaussian ensemble 
\begin{eqnarray}
  < l^\mu(t) > &=& 0 
  \label{Eq:noise-l}
\end{eqnarray}
and 
\begin{eqnarray}
  < l^\mu(t) l^\nu (t') > &=& 
  \frac{2 k_B T_N \chi_N }{\tau_{\rm sf}} 
  \delta_{\mu,\nu} \delta(t-t') 
  \label{Eq:noise-ll}, 
\end{eqnarray}
to satisfy the fluctuation-dissipation theorem~\cite{Kubo-text,Landau-text}. From now on we focus on the spin-wave region, where the magnetization ${\bm M}$ fluctuates only weakly around the ground state value $M_s {\bm \hat{\bm z}}$, and ${\bm M}/M_s={\bm \hat{\bm z}}+ {\bm m}$ is established to separate small fluctuations ${\bm m}$ from the ground state value. 

The central quantity that characterizes the spin Seebeck effect is the spin current $I_s$ injected into the nonmagnetic metal $N$, since it is proportional to the experimentally detectable electric voltage via the inverse spin Hall effect [Eq.~(\ref{Eq:ISHE02})]. This quantity can be calculated as the rate of change of the itinerant spin density in $N$ as $I_s = < \partial_ts^z(t) >$. Performing the perturbative approach in Eq.~(\ref{Eq:Bloch01}) in terms of $J_{\rm sd}$, we obtain 
\begin{equation}
  I_s(t)= \frac{J_{\rm sd}}{\hbar} \Im {\rm m} < s^+(t) m^-(t') >_{t' \to t}, 
\label{Eq:Is01} 
\end{equation}
where $s^\pm = s^x \pm \ui s^y$ and $m^\pm = m^x \pm \ui m^y$. Introducing the Fourier representation $f(t)=\int \frac{d \omega}{2 \pi} f_\omega e^{-\ui \omega t} $ and using the fact that the right hand side of Eq.~(\ref{Eq:Is01}) is only a function of $t-t'$ in the steady state, we obtain 
\begin{equation}
  I_s= \frac{J_{\rm sd}}{\hbar}  \int_{-\infty}^{\infty} 
\frac{d \omega}{2 \pi} \ll s^+_\omega m^-_{-\omega} \gg, 
\label{Eq:Is02} 
\end{equation}
where the average $\ll \cdots \gg$ is defined by $< s^+_\omega m^-_{\omega'} > = 2 \pi \delta (\omega + \omega') \ll s^+_\omega m^-_{-\omega} \gg $. 

To evaluate the right hand side of Eq.~(\ref{Eq:Is02}), the transverse components of Eqs.~(\ref{Eq:LLG01}) and (\ref{Eq:Bloch01}) are linearized with respect to $s^\pm$ and $m^\pm$. Then, to the lowest order in $J_{\rm sd}$, we obtain 
\begin{eqnarray}
  s^+_\omega &=& \frac{1}{-\ui \omega + \tau^{-1}_{\rm sf}} 
  \left( 
  l^+_\omega+   
  \frac{s_0 \tau^{-1}_{\rm sf}} 
       {\omega_0+\omega-\ui \alpha \omega} \gamma h^+_\omega \right) 
\end{eqnarray}
and 
\begin{eqnarray}
  m^-_\omega &=& \frac{1} 
  {\omega_0- \omega -\ui \alpha \omega} 
  \left( \gamma h^-_\omega+ 
    \frac{J_{\rm sd}}
    {-\ui \omega + \tau^{-1}_{\rm sf}}
    l^-_\omega 
  \right) , 
\end{eqnarray}
where $\omega_0= \gamma H_{0}$, $h^\pm= h^x \pm \ui h^y$, and $l^\pm= l^x \pm \ui l^y$. From the above equations, we see that ${\bm s}$ and ${\bm m}$ are affected by both the noise field ${\bm h}$ in $F$ and the noise source ${\bm l}$ in $N$ through the $s$-$d$ exchange interaction $J_{\rm sd}$ at the interface. Substituting the above equations into Eq.~(\ref{Eq:Is02}), the spin current injected into $N$ can be expressed as 
\begin{equation}
  I_s = I_s^{\rm pump} - I_s^{\rm back}, 
  \label{Eq:Is03}
\end{equation}
where $I_s^{\rm pump}$ and $I_s^{\rm back}$ are respectively defined by 
\begin{eqnarray}
I_s^{\rm pump} &=& -\frac{J_{\rm sd} s_0}{\hbar \tau_{\rm sf}} 
  \int_{-\infty}^\infty \frac{d \omega}{2 \pi} 
  \frac{\omega \ll \gamma h^+_\omega \gamma h^-_{-\omega} \gg }
       {|\omega-\omega_0+\ui \alpha \omega|^2 |\ui \omega -  \tau^{-1}_{\rm sf}|^2}
       \label{Eq:Is-pump01}
\end{eqnarray}
and 
\begin{eqnarray}
       I_s^{\rm back} &=& 
       -\frac{\alpha J_{\rm sd}^2 }{\hbar^2 } 
       \int_{-\infty}^\infty \frac{d \omega}{2 \pi} 
       \frac{\omega        \ll l^+_\omega l^-_{-\omega} \gg}
       {|\omega-\omega_0+\ui \alpha \omega|^2 |\ui \omega -  \tau^{-1}_{\rm sf}|^2}.  
       \label{Eq:Is-back01}
\end{eqnarray}

We readily see in this expression that $I_s^{\rm pump}$ represents the spin current pumped into $N$ by the thermal noise field ${\bm h}$ in $F$ (the so-called pumping component~\cite{Tserkovnyak05}), while $I_s^{\rm back}$ represents the spin current coming back into $F$ from the thermal noise source ${\bm l}$ in $N$ (the so-called backflow component~\cite{Foros05}). Using the two fluctuation-dissipation relations [Eqs.~(\ref{Eq:noise-hh}) and (\ref{Eq:noise-l})], the pumping and backflow components are finally calculated to be 
\begin{eqnarray}
I_s^{\rm pump} &=& - (G_s \kB/\hbar)T_F 
\label{Eq:Is-pump02}
\end{eqnarray}
and 
\begin{eqnarray}
I_s^{\rm back} &=& - (G_s \kB/\hbar)T_N, 
\label{Eq:Is-back02}
\end{eqnarray}
such that the net contribution can be summarized in the single expression 
\begin{equation}
  I_s = -G_s 
       \frac{\kB}{\hbar} \big( T_F- T_N \big), 
       \label{Eq:Is04} 
\end{equation}
where 
$G_s= -\frac{2 \alpha \tau^{-1}_{\rm sf} \chi_N J^2_{\rm sd}}{\hbar} 
  \int_{-\infty}^\infty \frac{d \omega}{2 \pi} 
  \left( \frac{\omega} 
       {|\omega-\omega_0+\ui \alpha \omega|^2 |\ui \omega -  \tau^{-1}_{\rm sf}|^2}
       \right) 
\approx 
J^2_{\rm sd} \chi_{N} \tau_{\rm sf}/\hbar $, 
and 
$a_S^3 M_s=\hbar \gamma $ is used. Here the negative sign before $G_s$ arises from defining the positive direction of $I_s$. Equation~(\ref{Eq:Is04}) is the basic equation to understand the spin Seebeck effect. 

At this stage it is important to note that, when the $z$ component of the quantity $<[ {\bm m} \times \partial_t {\bm m}]^z >$ is calculated from Eq.~(\ref{Eq:LLG01}) under the condition $\tau_{\rm sf}^{-1} \gg \omega_0$ and by neglecting the attachment of the nonmagnetic metal $N$, we can be show that the pumping component [Eq.~(\ref{Eq:Is-pump01})] can be expressed as 
\begin{equation}
I_s^{\rm pump} = 
- G_{\rm s} <[ {\bm m} \times \partial_t {\bm m}]^z >. 
\label{Eq:Is-pump03}
\end{equation}
On the other hand, from the above argument we observe that the backflow component is given by the same quantity evaluated at the {\it local thermal equilibrium}, i.e., 
\begin{equation}
I_s^{\rm back} = - G_{\rm s} 
<[ {\bm m} \times \partial_t {\bm m}]^z >_{\rm loc-eq}. 
\end{equation}
Therefore, the thermal spin injection by localized spins can alternatively be expressed as 
\begin{equation}
  I_s = -G_{\rm s} \Big( 
  < [ {\bm m} \times \partial_t {\bm m} ]^z >
  -  < [{\bm m} \times \partial_t 
    {\bm m} ]^z >_{\rm loc-eq}
  \Big). 
  \label{Eq:Is-pumpback01}
\end{equation}
This procedure was used in Ref.~\cite{Ohe11} to perform the numerical simulation of the spin Seebeck effect. 

Equations~(\ref{Eq:Is03}) and (\ref{Eq:Is04}) indicate that when both $F$ and $N$ are in local thermal equilibrium with a local equilibrium temperature $T_{l \mathchar`-eq}$ (i.e., $T_F=T_N=T_{l \mathchar`-eq}$), then there is no net spin injection into the attached nonmagnetic metal $N$. However, conversely, this means that if the ferromagnet $F$ deviates from the local thermal equilibrium for some reason, a finite spin current is injected into (or ejected from) the attached nonmagnetic metal $N$. Note that the local equilibrium temperature $T_{l \mathchar`-eq}$ is defined by the temperature of {\it optical} phonons having a localized nature with a large specific heat but small thermal conductivity, and that most of the phonon heat current is carried by {\it acoustic} phonons. Here it is important to point out that in a general nonequilibrium situation, each temperature $T_F$ or $T_N$ appearing in Eq.~(\ref{Eq:Is04}) should be identified as an effective magnon temperature $T^*_F$ or effective spin-accumulation temperature $T^*_N$ which characterizes the nonequilibrium state. One example of the definition of the effective temperature can be found in Ref.~\cite{Hohenberg89} where the distribution function of a nonequilibrium state is mimicked by a distribution function of an approximate equilibrium state with an effective temperature. In the subsequent sections we show that, even if there is no local equilibrium temperature difference between $F$ and $N$, effects of thermal diffusion of magnons or phonons in $F$ can generate a finite thermal spin injection into $N$, which can be regarded as a consequence of an effective temperature difference $T^*_F-T^*_N \neq 0$. 

These considerations lead to the following simple picture for the spin Seebeck effect. Namely, the essence of the spin Seebeck effect is that the localized spins in the ferromagnet are excited by the heat current flowing through the ferromagnet, which then generates finite spin injections because of the imbalance between the pumping component $I_s^{\rm pump}$ and backflow component $I_s^{\rm back}$. It is important to note that the heat current that excites the localized spins has two contributions: the magnon heat current and the phonon heat current. Accordingly, there are two relevant processes in the spin Seebeck effect. The first, in which the localized spins are excited by the magnon heat current, corresponds to the magnon-driven spin Seebeck effect discussed in Refs.~\cite{Xiao10} and~\cite{Adachi11}. The second, in which the localized spins are excited by the phonon heat current, corresponds to the phonon-drag spin Seebeck effect discussed in Ref.~\cite{Adachi10}.

\subsection{Linear-response approach to the magnon-driven spin Seebeck effect} 

\begin{figure}[t]
\begin{center}\scalebox{0.75}[0.75]{\includegraphics{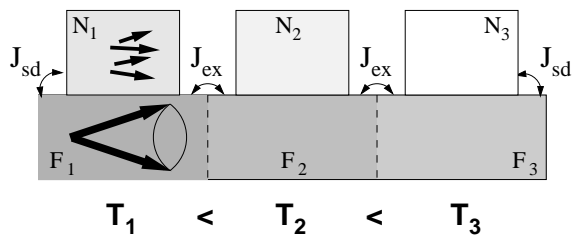}}\end{center} 
\caption{Schematic of system composed of ferromagnet $(F)$ and nonmagnetic metals $(N)$ divided into three temperature domains $F_1/N_1$, $F_2/N_2$, and $F_3/N_3$, with local temperatures $T_1$, $T_2$, $T_3$, respectively. 
Here, $J_{\rm sd}$ is the interface $s$-$d$ coupling between $F$ and $N$, and $J_{\rm ex}$ is the exchange coupling between two different segments in $F$. 
} 
\label{fig_SSEmodel01}
\end{figure}

In the previous section we gave the local picture of thermal spin injection by magnons, but we did not discuss effects of magnon diffusion. Now, starting from the equivalent model to the previous subsection, we first reformulate the thermal spin injection by magnons in terms of the quantum many-body theory, and then extend it to thermal spin injection containing the effect of magnon diffusion~\cite{Adachi11}. Let us consider the model shown in Fig.~\ref{fig_SSEmodel01} where the ferromagnet and the attached nonmagnetic metals are interacting weakly through interface $s$-$d$ exchange coupling $J_{\rm sd}$. This model is essentially the same as that considered in the previous subsection (Fig.~\ref{fig_SpinPump02}). An important point in this model is that there is no {\it local} temperature difference between the ferromagnet and the attached nonmagnetic metals, i.e., $T_{N_1}=T_{F_1}=T_1$, $T_{N_2}=T_{F_2}=T_2$, and $T_{N_3}=T_{F_3}=T_3$. It is assumed that each domain is initially in local thermal equilibrium without interactions with the neighboring domains. We then switched on the interactions between the domains, and calculate the nonequilibrium dynamics of the system.

The localized spin in the ferromagnet is described by the exchange Hamiltonian 
\begin{eqnarray}
  {\cal H}_{\rm ex} &=& - J_{\rm ex} \sum_{\langle \bmr_i, \bmr_j \rangle } \; 
  {\bm S}(\bmr_i) \cdot {\bm S}(\bmr_j) 
  - \sum_{\bmr_i}
  \gamma \hbar {\bm H}_0 \cdot {\bm S}(\bmr_i), 
  \label{Eq:H_ex00}
\end{eqnarray}
where $\langle \bmr_i, \bmr_j \rangle $ denotes a pair of the nearest neighbors. In addition to Eq.~(\ref{Eq:H_ex00}), we later consider a selfenergy correction to represent the Gilbert damping term in the magnon propagator. The single-particle Hamiltonian for conduction electrons in the nonmagnetic metal is given by 
\begin{eqnarray}
  {\cal H}_{N} &=& \sum_{\bmp,\bmp' } 
  c^\dag_\bmp \Big\{ \epsilon_\bmp \delta_{\bmp,\bmp'} 
  + U_{\bmp-\bmp'} [1+ \ui \eta_{\rm so} 
  {\bm \sigma}\cdot (\bmp \times \bmp')] \Big\} c_{\bmp'}, 
  \label{Eq:H_el01}
\end{eqnarray}
where $c^\dag_\bmp=(c^\dag_{\bmp,\uparrow},c^\dag_{\bmp,\downarrow})$ is the electron creation operator for spin projection $\uparrow$ and $\downarrow$, $U_{\bmp-\bmp'}$ is the Fourier transform of the impurity potential $U_{\rm imp} \sum_{\bmr_0 \in {\rm impurities}} \delta(\bmr-\bmr_0)$, and $\eta_{so}$ is the strength of the spin-orbit interaction~\cite{Takahashi08}. Finally at the $F$-$N$ interface, the magnetic interaction between the conduction-electron spin density and localized spin is described by the $s$-$d$ Hamiltonian, 
\begin{eqnarray}
  {\cal H}_{F \mathchar`- N} &=& 
  \frac{1}{\sqrt{N_F N_N}} \sum_{\bmq,\bmk} 
       {\cal J}^{\bmk-\bmq}_{\rm sd}{\bm s}_\bmk \cdot {\bm S}_\bmq, 
\end{eqnarray}
where 
$\bms_\bmk = \frac{1}{\sqrt{N_N}} 
\sum_{\bmp} c^\dag_{\bmp+\bmk} {\bm \sigma}c_{\bmp}$ 
is the spin-density operator of conduction electrons, 
$\bmS_\bmq= \frac{1}{\sqrt{N_F}} \sum_{\bmq} \bmS(\bmr_i) e^{\ui \bmq \cdot \bmr}$ 
is the localized spin operator at the interface, and $N_F$ ($N_N$) is the number of lattice sites in $F$ ($N$) in each domain. Here, ${\cal J}^{\bmk-\bmq}_{\rm sd}$ is the Fourier transform of 
${\cal J}_{\rm sd} (\bmr) = J_{\rm sd} \sum_{{\bm r}_0 \in {\rm F-N \, interface}} 
a_S^3 \delta({\bm r}-{\bm r}_0)$ 
with $J_{\rm sd}$ being the strength of the $s$-$d$ exchange interaction. 

The spin current induced in the nonmagnetic metal $N_i$ ($i=1,2,3$) can be calculated as the rate of change of the spin accumulation in $N_i$, i.e., 
$I_s(t) \equiv 
\sum_{\bmr \in N_i} \langle \partial_t s^z({\bm r}, t) \rangle 
= \langle \partial_t \widetilde{s}^z_{\bmk_0}(t) \rangle_{\bmk_0 \to {\bm 0}} $, 
where $\langle \cdots \rangle$ means the statistical average at a given time $t$, and $\widetilde{\bms}_\bmk= \sqrt{N_N} \bms_\bmk$. Introducing the magnon operator 
\begin{eqnarray}
S^x(\bmr_i) &=& \sqrt{ \frac{S_0}{2 N_{\rm F}} } \sum_{\bmq} 
(a^\dag_{-\bmq}+ a_\bmq )e^{\ui \bmq \cdot \bmr_i}, 
\label{Eq:magnon-Sx01} \\
S^y(\bmr_i) &=&  -\ui \sqrt{ \frac{S_0}{2 N_{\rm F}} } 
\sum_{\bmq} (a^\dag_{-\bmq}- a_\bmq)e^{\ui \bmq \cdot \bmr_i}, 
\end{eqnarray}
and 
\begin{eqnarray} 
\label{Eq:magnon-Sy01} \\
S^z(\bmr_i) &=& -S_0+  \frac{1}{N_{\rm F}} \sum_{\bmq,\bmQ} 
a^\dag_{\bmq} a_{\bmq+\bmQ} e^{\ui \bmQ \cdot \bmr_i}, 
\label{Eq:magnon-Sz01}
\end{eqnarray}
the Heisenberg equation of motion for $\widetilde{s^z}_{\bmk_0}$ yields 
\begin{eqnarray} 
  \partial_t \widetilde{s}^z_{\bmk_0} 
  &=& 
  \ui \sum_{\bmq,\bmk} \frac{2{\cal J}^{\bmk-\bmq}_{\rm sd} \sqrt{S_0} }
      {\sqrt{2 N_F N_N} \hbar } 
      \Big( a^+_\bmq s^-_{\bmk+\bmk_0} 
      - a^-_\bmq s^+_{\bmk+\bmk_0} \Big), 
\end{eqnarray} 
where $S_0$ is the size of the localized spins in $F$. Here have used the relation 
$[\widetilde{s}^z_{\bmk},\widetilde{s}^\pm_{\bmk'}] 
= \pm 2 \widetilde{s}^\pm_{\bmk+\bmk'} $ 
and neglected a small correction term arising from the spin-orbit interaction, assuming that the spin-orbit interaction is weak enough in the vicinity of the interface. The statistical average of the above quantity gives the spin current 
\begin{eqnarray}
  I_s(t) &=&   \sum_{\bmq,\bmk} 
  \frac{-4{\cal J}^{\bmk-\bmq}_{\rm sd} \sqrt{S_0} }
  {\sqrt{2 N_F N_N} \hbar} 
  \Re {\rm e} C^{<}_{\bmk,\bmq}(t,t) 
  \label{Eq:Is-diag01}, 
\end{eqnarray}
where 
$C^{<}_{\bmk,\bmq}(t,t') = - \ui \langle  a^+_\bmq(t') s^-_\bmk(t) \rangle $ 
measures the correlation between the magnon operator $a_\bmq^+$ and the spin-density operator $s^-_\bmk= (s^x_\bmk- \ui s^y_\bmk)/2$. Note that the time dependence of $I_s(t)$ vanishes in the steady state and hence is hereafter discarded. Introducing the frequency representation 
$C^{<}_{\bmk,\bmq}(t-t') = 
\int_{-\infty}^\infty \frac{d \omega}{2 \pi} 
{C}^{<}_{\bmq,\bmk}(\omega) e^{- \ui \omega (t-t')}$, 
adopting the representation~\cite{Larkin75} 
$\check{C} = \left({{C^{R}, C^{K}} \atop {0 \;\;\;  ,C^{A}}} \right)$, 
and using the relation $C^{<}= \frac{1}{2} [C^{K}- C^{R} + C^{A}]$, we obtain 
\begin{equation}
  I_s =   \sum_{\bmq,\bmk} 
  \frac{-2{\cal J}^{\bmk-\bmq}_{\rm sd} \sqrt{S_0} }
  {\sqrt{2 N_F N_N} \hbar} 
  \Re {\rm e} \int_\omega C^{K}_{\bmk,\bmq}(\omega) 
  \label{Eq:I_s01}
\end{equation}
for the thermal spin current $I_s$ in the steady state, where we have introduced the shorthand notation $\int_\omega= \int_{-\infty}^\infty \frac{d \omega}{2 \pi}$. 

\begin{figure}[t]
\begin{center} \scalebox{0.75}[0.75]{\includegraphics{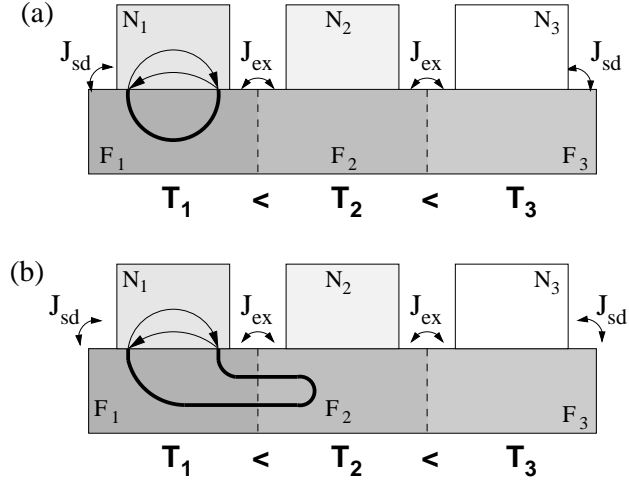}} \end{center} 
\caption{Feynman diagrams expressing the magnon-driven spin Seebeck effect~\cite{Adachi11}. The thin solid lines with arrows (bold lines without arrows) represent electron propagators (magnon propagators). } 
\label{fig_SSEdiag01}
\end{figure}

We first consider the process shown in Fig.~\ref{fig_SSEdiag01} (a) where magnons travel around the ferromagnet $F_1$ without sensing the temperature difference between $F_1$ and $F_2$. Using the standard rules of constructing the Feynman diagram in Keldysh space~\cite{Rammer86}, the corresponding interface Green's function $\check{C}_{\bmk,\bmq}(\omega)$ for the correlation between the magnons in $F_1$ and the itinerant spin density in $N_1$ can be written in the form 
\begin{eqnarray}
  \check{C}_{\bmk,\bmq} (\omega) &=& 
  \frac{{\cal J}^{\bmk-\bmq}_{\rm sd} \sqrt{S_0} }{\sqrt{N_N N_F}}
  \check{\chi}_\bmk(\omega) \check{X}_\bmq(\omega), 
  \label{Eq:C-func01} 
\end{eqnarray}
where $N_N$ and $N_F$ are the number of lattice sites in $N_1$ and $F_1$. In the above equation, $\check{\chi}_{\bmk}(\omega)$ is the spin-density propagator 
\begin{eqnarray}
\check{\chi}_{\bmk}(\omega) &=& 
\left( { \chi^R_\bmk (\omega), \atop 0,} 
     { \chi^K_\bmk (\omega)\atop \chi^A_\bmk (\omega)} \right), 
\end{eqnarray}
while $\check{X}_{\bmq}(\omega)$ is the bare magnon propagator 
\begin{eqnarray}
\check{X}_{\bmq}(\omega) &=&
\left( { X^R_\bmq (\omega), \atop 0,} 
     { X^K_\bmq (\omega)\atop X^A_\bmq (\omega)} \right), 
\end{eqnarray}
both of which satisfy the equilibrium condition: 
\begin{eqnarray}
\chi^A_\bmk(\omega)= [\chi^R_\bmk(\omega)]^*,&&
\chi^K_\bmk(\omega)= 2 \ui \, {\rm Im} \chi^R_\bmk(\omega)
\coth(\frac{\hbar \omega}{2 \kB T} ) 
\label{Eq:chi-eq01} 
\end{eqnarray} 
and 
\begin{eqnarray}
X^A_\bmq(\omega)= [X^R_\bmq(\omega)]^*,&&
X^K_\bmq(\omega)= 2 \ui \, {\rm Im} X^R_\bmq(\omega) 
\coth(\frac{\hbar \omega}{2 \kB T}). 
\label{Eq:X-eq01} 
\end{eqnarray} 
The retarded component of $\check{\chi}_{\bmk}(\omega)$ is given by 
$\chi^R_\bmk(\omega)= \chi_N/(1+ \lambda_N^2 k^2 - \ui \omega \tau_{\rm sf})$ ~\cite{Fulde68} 
where $\chi_N$, $\lambda_N$, and $\tau_{\rm sf}$ are the paramagnetic susceptibility, the spin diffusion length, and spin relaxation time. The retarded component of $\check{X}_{\bmq}(\omega)$ is given by $X^R_\bmq(\omega)= (\omega- \omega_\bmq + \ui \alpha \omega)^{-1}$, where $\alpha$ is the Gilbert damping constant and $\omega_\bmq= \gamma H_0 + D_{\rm ex} q^2$ is the magnon frequency with $D_{\rm ex}$ being the exchange stiffness. 

Substituting Eq.~(\ref{Eq:C-func01}) into Eq.~(\ref{Eq:I_s01}) and using the equilibrium conditions [Eqs.~(\ref{Eq:chi-eq01}) and ~(\ref{Eq:X-eq01})], we obtain the expression for the spin current injected into $N_1$ 
\begin{eqnarray}
  I_s &=& 
  -\frac{4 N_{\rm int} {J}_{\rm sd}^2 S_0}{\sqrt{2}\hbar^2 N_{N} N_{F}} 
  \sum_{\bmq,\bmk} 
  \int_{\omega} 
       {\rm Im} \chi_{\bmk}^{R}(\omega) 
       {\rm Im} X_{\bmq}^{R}(\omega) \nonumber \\
    && \qquad \times 
       \left[ \coth(\frac{\hbar \omega}{2 \kB T_{N_1}}) 
         - \coth(\frac{\hbar \omega}{ 2 \kB T_{{F_1}}})   \right], 
      \label{Eq:Is_local01} 
\end{eqnarray}
where $N_{\rm int}$ is the number of localized spins at the interface. From Eq.~(\ref{Eq:Is_local01}), it is clear that no spin current is injected into the nonmagnetic metal $N_1$ when $N_1$ and $F_1$ have the same temperature. 

The above result that the injected spin current vanishes when $T_{F_1}=T_{N_1}$ originates from the equilibrium condition of the magnon propagator [Eq.~(\ref{Eq:X-eq01})]. When the magnons deviate from local thermal equilibrium, the magnon propagator cannot be written in the equilibrium form, and it generates a new contribution. To see this, let us consider the process shown in Fig.~\ref{fig_SSEdiag01} (b) where the magnons feel the temperature difference between $F_1$ and $F_2$ through the following magnon interaction between $F_1$ and $F_2$: 
\begin{eqnarray}
  {\cal H}_{F \mathchar`- F} &=& 
  -\frac{1}{N_F} \sum_{\bmq,\bmq'} 2 {\cal J}_{\rm ex}^{\bmq - \bmq'} 
  {\bm S}_\bmq \cdot {\bm S}_{-\bmq'}, 
\end{eqnarray}
where ${\cal J}_{\rm ex}^{\bmq-\bmq'}$ is the Fourier transform of 
${\cal J}_{\rm ex}(\bmr) = 
J_{\rm ex} \sum_{\bmr_0 \in F \mathchar`- F \; {\rm interface}} 
a_S^3 \delta(\bmr-\bmr_0)$. 
We now treat all of the magnon lines as a single magnon propagator $\delta \check{X}_\bmq(\omega)$ in the following way: 
\begin{eqnarray}
\delta \check{X}_\bmq(\omega) &=& 
\frac{1}{N_F^2}\sum_{\bmq'} |{\cal J}_{\rm ex}^{\bmq-\bmq'}|^2 
\check{X}_\bmq (\omega) \check{X}_{\bmq'} (\omega) \check{X}_\bmq (\omega). 
\end{eqnarray}
Then the propagator is decomposed into the local-equilibrium part and nonequilibrium part via~\cite{Michaeli09} 
\begin{eqnarray} 
  \delta \check{X}_\bmq (\omega) &=& 
  \delta \check{X}^{l \mathchar`-eq}_\bmq (\omega) 
  + \delta \check{X}^{n \mathchar`-eq}_\bmq (\omega), 
  \label{Eq:Xnonloc01}   
\end{eqnarray}
where 
\begin{eqnarray}
\delta \check{X}^{l \mathchar`-eq}_{\bmq}(\omega) &=&
\left( { \delta X^{l \mathchar`-eq,R} (\omega), \atop 0,} 
     { \delta X^{l \mathchar`-eq,K} (\omega) 
       \atop \delta X^{l \mathchar`-eq,A} (\omega)} 
     \right) 
     \label{Eq:dX_loceq01}
\end{eqnarray}
is the local-equilibrium propagator satisfying the local-equilibrium condition, i.e., 
$\delta X^{l \mathchar`-eq,A}_\bmq = 
[\delta X^{l \mathchar`-eq,R}_\bmq]^* $ 
and 
$\delta X^{l \mathchar`-eq,K}_\bmq= 
[\delta X^{l \mathchar`-eq,R}_\bmq- \delta X^{l \mathchar`-eq,A}_\bmq] 
\coth(\frac{\hbar \omega}{2 \kB T} ) $ 
with 
\begin{eqnarray}
  \delta X^{l \mathchar`-eq,R}_\bmq(\omega) &=& 
  \frac{1}{N_F^2} \sum_{\bmq'} |{\cal J}_{\rm ex}^{\bmq-\bmq'}|^2
  \Big( X^R_\bmq(\omega) \Big)^2 X^R_{\bmq'}(\omega), 
\end{eqnarray}
while 
\begin{eqnarray}
\delta \check{X}^{n \mathchar`-eq}_{\bmq}(\omega) &=&
\left( { 0, \atop 0,} 
     { \delta X^{n \mathchar`-eq,K} (\omega) \atop 0} \right) 
     \label{Eq:dX_noneq01}
\end{eqnarray}
is the nonequilibrium propagator with 
$\delta {X}_\bmq^{n \mathchar`-eq,K} (\omega)$ 
given by 
\begin{eqnarray}
  \delta {X}_\bmq^{n \mathchar`-eq,K} (\omega) &=& 
  \sum_{\bmq'}  \frac{|2{\cal J}_{\rm ex}^{\bmq-\bmq'}S_0|^2}{N_F^2} 
  \Big[ {X}^R_{\bmq'}(\omega)-{X}^A_{\bmq'}(\omega) \Big] \nonumber \\ 
    &\times & 
  |{X}^R_\bmq(\omega)|^2  
  \big[ \coth(\frac{\hbar \omega}{2 \kB T_{F_2}}) 
    - \coth(\frac{\hbar \omega}{2 \kB T_{{F_1}}}) \big] . 
  \label{Eq:dX_noneqK01}
\end{eqnarray}

When we substitute Eq.~(\ref{Eq:Xnonloc01}) into Eq.~(\ref{Eq:I_s01}) and use Eq.~(\ref{Eq:C-func01}) with $\check{X}_\bmq(\omega)$ replaced by $\delta \check{X}_\bmq(\omega)$, we obtain the following expression for the magnon-mediated thermal spin injection: 
\begin{eqnarray}
  I_s &=&  \frac{-4J^2_{\rm sd}S_0 (2 J_{\rm ex} S_0)^2 N_{\rm int} N'_{\rm int} }
  {\sqrt{2} \hbar^2 N_F^3 N_N } 
  \sum_{\bmq,\bmq',\bmk} 
  \int_\omega 
      {\rm Im} \chi_{\bmk}^R(\omega) 
      \nonumber \\
      &\times& 
      |X_{\bmq}^{R}(\omega)|^2 
      {\rm Im} X_{\bmq'}^R(\omega)  
      [ \coth(\frac{\hbar \omega}{2 \kB T_{1}}) 
        - \coth(\frac{\hbar \omega}{2 \kB T_{2}}) ], 
\end{eqnarray}
where $N'_{\rm int}$ is the number of localized spins at the $F_1 \mathchar`- F_2$ interface. The integration over $\omega$ can be performed by picking up only magnon poles under the condition $\alpha \hbar \omega_\bmq \ll \kB T_{F_1}, \kB T_{N_1}$ (which is always satisfied when magnon excitations are well defined), yielding  
$
\int_\omega
{\rm Im} \chi_{\bmk}^R(\omega) 
|X_{\bmq}^R(\omega)|^2 
{\rm Im} X_{\bmq'}^R(\omega)  
[ \coth(\frac{\hbar \omega}{2 \kB T_{1}})- \coth(\frac{\hbar \omega}{2 \kB T_{2}}) ] 
\approx 
\frac{- \pi }
{2 \alpha \widetilde{\omega}_{\bmq}} 
\delta(\omega_{\bmq}-\omega_{\bmq'}) 
{\rm Im}\chi_{\bmk}^R(\widetilde{\omega}_\bmq) 
[ \coth(\frac{\hbar \widetilde{\omega}_\bmq}{2 \kB T_{1}}) 
- \coth(\frac{\hbar \widetilde{\omega}_\bmq}{2 \kB T_{2}}) ] 
$. 
With the classical approximation 
$\coth(\frac{ \hbar \widetilde{\omega}_\bmq}{2 \kB T}) \approx \frac{2 \kB T}{\hbar \widetilde{\omega}_\bmq}$, 
we obtain 
\begin{eqnarray}
  I_s &=&  
  \frac{N_{\rm int} (J^2_{\rm sd} S_0) \chi_N \tau_{\rm sf} (a/\lambda_N)^3} 
       {8 \sqrt{2} \pi^5 \hbar^3 \alpha (\Lambda/a_S)} 
       \Upsilon_2 \kB (T_1- T_2), 
      \label{Eq:Is_nonlocal03} 
\end{eqnarray}
where $\Lambda$ is the size of $F_1$ along the temperature gradient, and 
$\Upsilon_2 = \int_0^1 d x \int_0^1 d y \frac{y^2} {[(1+x^2)^2+ y^2 (2 S_0 J_{\rm ex} \tau_{\rm sf}/\hbar)^2  ]}$ 
which is approximated as $\Upsilon_2 \approx 0.1426$ ($\Upsilon_2 \approx 0.337 \hbar/ 2 S_0 J_{\rm ex} \tau_{\rm sf}$) for $2 S_0 J_{\rm ex} \tau_{\rm sf}/\hbar {_< \atop ^\sim} 1$ (for $2 S_0 J_{\rm ex}\tau_{\rm sf}/\hbar \gg 1$). Eq.~(\ref{Eq:Is_nonlocal03}) expresses the signal of the magnon-driven spin Seebeck effect.

The result obtained above can be understood in the following way. 
In this process there is no vertical temperature difference between $F_1$ and $N_1$ (i.e., $T_{F_1}= T_{N_1}$), and hence a naive use of Eq.~(\ref{Eq:Is04}) cannot explain the result. However, as pointed out in the last part of the previous section, each temperature $T_F$ or $T_N$ in Eq.~(\ref{Eq:Is04}) should be understood as an effective magnon temperature $T^*_F$ or effective spin-accumulation temperature $T^*_N$. 
In the present situation, because there is a horizontal temperature difference $T_1- T_2$, the magnon heat current flows in the horizontal direction. This heat current brings about a deviation of the effective magnon temperature $T^*_{F_1}$ from $T_{F_1}$ (i.e., $T^*_{F_1} \neq T_{F_1}$), whereas the effective spin-accumulation temperature $T^*_N$ remains unchanged (i.e., $T^*_{N_1}=T_{N_1}$) because the nonmagnetic metal $N_1$ is isolated and not extended in the horizontal direction. 
Therefore, the resultant effective temperature difference ($T^*_{F_1} -T^*_{N_1} \neq 0$) drives the thermal spin injection in accordance with Eq.~(\ref{Eq:Is04}).

\subsection{Length scale associated with the spin Seebeck effect} 

We have already seen in Eq.~(\ref{Eq:Is-pump03}) that the pumping component is given by the quantity $I_s^{\rm pump}= -G_s <[ {\bm m} \times \partial_t {\bm m}]^z >$. Using this result and the scenario of the magnon-driven spin Seebeck effect, let us calculate the spatial dependence of $I_s^{\rm pump}(\bmR)$ and discuss the length scale associated with the spin Seebeck effect. The starting point is the Landau-Lifshitz-Gilbert equation for a bulk ferromagnet written in the form~\cite{Barnes-Maekawa05} 
\begin{eqnarray}
  \partial_t {\bm M}(\bmR,t) &=& 
  - {\bm \nabla} \cdot {\bm J}^{\bm M} (\bmR,t) \nonumber \\
  &&+ {\bm M} (\bmR,t)  \times 
       \Big( 
       -\gamma [ {\bm H}_{0} + {\bm h}(\bmR,t) ] 
      + \frac{\widehat{\alpha}}{M_s} \partial_t {\bm M}(\bmR,t)   \Big), 
  \label{Eq:LLG02} 
\end{eqnarray}
where the $M^{\mu}$ component of the magnetization current ${\bm J}^{\bm M}$ is given by~\cite{Lifshitz-Pitaevskii} 
\begin{equation}
  J_j^{M^{\mu}} = \frac{D_{\rm ex}}{\hbar M_s} [{\bm M} \times \nabla_{j} {\bm M}]^\mu 
\end{equation}
with $D_{\rm ex}$ being the exchange stiffness. Here the Greek indices refer to the components in spin space, and the Latin indices refer to the components in the real space. In Eq.~(\ref{Eq:LLG02}) the Gilbert damping factor $\widehat{\alpha}$ is an anisotropic tensor~\cite{Safonov02} to account for the difference between the transverse dynamics and longitudinal dynamics~\cite{Mori62}, and it is represented here as $\widehat{\alpha}= {\rm diag}(\alpha_\perp,\alpha_\perp,\alpha_\parallel)$. Note that the transverse damping $\alpha_\perp$ is relevant to the ferromagnetic resonance experiment, while information on the longitudinal damping $\alpha_\parallel$ is quite difficult to obtain from experiments. As before, the thermal noise field is given by the Gaussian white noise obeying 
\begin{eqnarray}
< h^\mu(\bmR_i,t) > &=& 0 
\label{Eq:noise-h02}
\end{eqnarray}
and 
\begin{eqnarray}
< h^\mu(\bmR_i, t) h^\nu (\bmR_j, t')> &=& 
\frac{2 k_B T(\bmR_i) \alpha_{\mu,\nu} }{\gamma a_S^3 M_s} 
\delta_{\mu,\nu} \delta_{ij} \delta(t-t'), 
\label{Eq:noise-hh02} 
\end{eqnarray}
where $\alpha_{\mu,\nu}= \alpha_\parallel$ for $\mu=\nu=x,y$ and $\alpha_{\mu,\nu}= \alpha_\perp$ for $\mu=\nu=z$~\cite{Kubo70}. We use again the spin-wave approximation ${\bm M}/M_s={\bm \hat{\bm z}}+ {\bm m}$ and rotating-frame representation $m^\pm = m^x \pm \ui m^y$. Using the transverse component of the Landau-Lifshitz-Gilbert equation~(\ref{Eq:LLG02}) and taking its statistical average, the pumping current $I_s^{\rm pump}(\bmR)= - G_s <m^x(\bmR,t) \partial_t m^y(\bmR,t) -m^y(\bmR,t) \partial_t m^x(\bmR,t)>$ is calculated to be 
\begin{eqnarray}
I_s^{\rm pump}(\bmR) &=&
-\frac{G_s}{2}  
\Big( \omega(-\ui \nabla_{\bmr_1}) + \omega(-\ui \nabla_{\bmr_2}) \Big) \nonumber \\
&& \times < m^+ (\bmr_1,t)  m^- (\bmr_2,t)>_{\bmr_1,\bmr_2 \to \bmR} 
\end{eqnarray}
where $\omega(-\ui \nabla)= \gamma H_{0} + D_{\rm ex} (-\ui \nabla)^2$. 
Because the above equation contains the gradient operator $\nabla_\bmr$ acting solely on one of the pairs in the correlator $<m^+ (\bmr_1,t) m^- (\bmr_2,t)>$, it is useful to introduce the Wigner representation with respect to the spatial coordinate in the following manner~\cite{Schopohl80,Serene83}: 
\begin{equation}
  \bmR = \frac{1}{2} ( \bmr_1+ \bmr_2 ), \qquad 
  \bmr = \bmr_1- \bmr_2, 
  \label{Eq:Wigner01}
\end{equation}
where $\bmR$ represents the center of mass coordinate, while $\bmr$ represents the relative coordinate. In this representation, we have the relation 
\begin{eqnarray}
<m^+ (\bmr_1,t) m^- (\bmr_2,t)>_{\bmr_1,\bmr_2 \to \bmR} 
&=& 
-2 \sum_\bmq < m^z_\bmq(\bmR,t)> e^{\ui \bmq \cdot \bmr} \Big|_{\bmr \to {\bm 0}}, 
\end{eqnarray}
where 
$< m^z_\bmq(\bmR,t)>= 
-\frac{1}{2N} \sum_{\bmK} < m^+_{\bmq+\bmK/2}(t) m^-_{\bmq-\bmK/2}(t)> 
e^{\ui \bmK \cdot \bmR}$, 
and we have introduced the Fourier transformation 
$m^-(\bmr)= \frac{1}{\sqrt{N}} \sum_\bmk m^-_\bmk e^{\ui \bmk \cdot \bmr}$. 
This allows us to represent the pumping current as 
\begin{equation}
I_s^{\rm pump}(\bmR) 
=   2 G_s \sum_\bmq \omega_\bmq < m^z_\bmq(\bmR,t)>, 
\label{Eq:Is-pump04}
\end{equation}
where $\omega_\bmq= \omega_0+ D_{\rm ex} q^2$, and we have used the quasiclassical approximation $|\bmK| \ll |\bmq| $. 

To calculate the pumping current from Eq.~(\ref{Eq:Is-pump04}), we take the statistical average of the $z$ component of the LLG equation~(\ref{Eq:LLG02}): 
\begin{eqnarray}
  \partial_t < m^z(\bmR,t)> 
  &=& 
  - \nabla_\bmR \cdot < \bmJ^{m^z}(\bmR,t) > 
  - 2 \alpha_\parallel \sum_\bmq \omega_\bmq  < m^z_\bmq(\bmR,t)> \nonumber \\
  && + \Im {\rm m} < m^+ (\bmR,t) \gamma h^- (\bmR,t)>, 
 \label{Eq:LLGz01}
\end{eqnarray}
where the last term is evaluated with the help of the Wigner representation (\ref{Eq:Wigner01}) to give 
\begin{equation}
  \Im {\rm m} < m^+ (\bmR,t) \gamma h^- (\bmR,t)> 
  = - \frac{2 \alpha_\parallel \kB T(\bmR) \gamma}{a_S^3 M_s}, 
\end{equation}
where we have used the Fourier representation in frequency space in the intermediate step of the calculation.

We use the following assumptions to solve Eq.~(\ref{Eq:LLGz01}) in a closed form. First, we assume Fick's law of magnon diffusion, 
\begin{equation}
  < \bmJ^{m^z}(\bmR,t) > = - {\cal D} \nabla_\bmR < m^z (\bmR,t)> , 
  \label{Eq:LLGz-approx01}  
\end{equation}
where ${\cal D}$ is the diffusion constant. Second, we introduce a wavenumber $\bmq_0$ roughly corresponding to the thermal de Broglie wavenumber with kinetic energy $\kB T$~\cite{Reichl-text}, which satisfies 
\begin{equation}
  \sum_\bmq \omega_\bmq  < m^z_\bmq(\bmR,t)> 
  \approx \omega_{\bmq_0} < m^z (\bmR,t)>  
  \label{Eq:LLGz-approx02}
\end{equation}
where we have used $ < m^z(\bmR,t)> = \sum_\bmq  < m^z_\bmq(\bmR,t)> $. Substituting Eqs.~(\ref{Eq:LLGz-approx01}) and~(\ref{Eq:LLGz-approx02}) into Eq.~(\ref{Eq:LLGz01}), we obtain 
\begin{equation}
  \Big( \partial_t   - {\cal D}\nabla_\bmR^2  \Big) < m^z(\bmR,t)> 
  = 
  -2 \alpha_\parallel \omega_{\bmq_0} < m^z(\bmR,t)>  
  - \frac{2 \alpha_\parallel \kB T(\bmR) \gamma}{a_S^3 M_s},  
  \label{Eq:LLGz02}
\end{equation}
where the right hand side represents the sink due to the longitudinal Gilbert damping (the first term) and source due to the heat bath (the second term). This equation can be solved in terms of the magnon distribution $< m^z(\bmR,t)>$. 

Now we evaluate the spatial dependence of the spin Seebeck effect. From Eq.~(\ref{Eq:Is-pump04}), thermal spin injection by localized spins is given by 
\begin{equation}
I_s (\bmR) 
=   2 G_s \omega_{\bmq_0} \Big( < m^z (\bmR,t)> 
- < m^z(\bmR,t)>_{\rm loc-eq} \Big), 
\label{Eq:Is-pump05}
\end{equation}
where we have considered the contribution from the backflow component [Eq.~(\ref{Eq:Is-pumpback01})] and used the approximation [Eq.~(\ref{Eq:LLGz-approx02})]. Under the local equilibrium condition there is no magnon diffusion, and by setting the both sides of Eq.~(\ref{Eq:LLGz02}) equal to zero, we calculate the local equilibrium magnon distribution to be 
\begin{equation}
  < m^z(\bmR,t)>_{\rm loc-eq} 
  =
  -\frac{\kB T(\bmR)}{\hbar \omega_{\bmq_0}}, 
\end{equation}
where we have used $a_S^3 M_s = \gamma \hbar$. This equation represents the classical limit of the magnon distribution function $(e^{\hbar \omega_{\bmq_0}/\kB T}-1)^{-1}$ as it should because we neglect the quantum fluctuation in the fluctuation-dissipation relation~(\ref{Eq:noise-hh02}). In a current-carrying steady state with magnon diffusion, we can set the time derivative equal to zero in Eq.~(\ref{Eq:LLGz02}), and by putting 
$<m^z(\bmR,t)> - < m^z(\bmR,t)>_{\rm loc-eq} = < \delta m^z(\bmR,t)> $ 
we obtain 
\begin{equation}
  \nabla^2_\bmR <\delta m^z (\bmR,t) > = \frac{1}{\lambda_{\rm m}^2} <\delta m^z (\bmR,t) >, 
\end{equation}
where we have introduced a new length 
\begin{equation}
\lambda_{\rm m}^2 = {\cal D}/(2 \alpha_\parallel \omega_{\bmq_0}). 
\label{Eq:lambda_m01}
\end{equation}

As is clear from the fact that the thermal spin injection by localized spins is given by $<\delta m^z (\bmR,t) >$ [see Eq.~(\ref{Eq:Is-pump05})], $\lambda_{\rm m}$ corresponds to the length scale associated with the magnon-driven spin Seebeck effect. Physically, $\lambda_{\rm m}$ corresponds to the length associated with magnon number conservation, or in other words it is an energy relaxation length for magnons. In the case of the phonon-drag spin Seebeck effect, $\lambda_{\rm m}$ is replaced by $\lambda_{\rm p}$, which corresponds to the length associated with phonon number conservation, or in other words it is an energy relaxation length of phonons. 

For the scenario of the magnon-driven spin Seebeck effect to be valid, the length scale given by Eq.~(\ref{Eq:lambda_m01}) should be as long as a millimeter because such a long length scale is observed in experiments~\cite{Uchida08,Jaworski10,Uchida10a} (see Fig.~\ref{fig:transverseSSE2}). However, as we have already noted, experimental information on the {\it longitudinal} damping constant $\alpha_\parallel$ is lacking, such that no reliable estimate of $\lambda_{\rm m}$ is available at the moment. This is because, while the damping $\alpha_\parallel$ roughly corresponds to the longitudinal relaxation time $T_1$ in the case of nuclear magnetic resonance, the longitudinal relaxation in the ferromagnetic resonance is not well defined. An experiment detecting the propagation of a wavepacket of {\it exchange magnons}, not magnetostatic magnons, may be able to estimate the magnitude of $\lambda_{\rm m}$.

\section{Phonon-Drag Contribution to the Spin Seebeck Effect} 

Phonon drag is a well-established idea in thermoelectricity~\cite{Blatt76,Physical-Kinetics}. Back in 1946, in the context of thermoelectricity, Gurevich pointed out that thermopower can be generated by nonequilibrium phonons driven by a temperature gradient, which then drag electrons and cause their motions~\cite{Gurevich46}. This idea, now known as phonon drag, has been established as the principal mechanism behind low-temperature enhancement of thermopower. Here, nonequilibrium phonons are the key. In this section, we first discuss acoustic spin pumping to understand the role of nonequilibrium phonons in the spin Seebeck effect. Then we present a microscopic approach to the phonon-drag contribution to the spin Seebeck effect. 

\begin{figure}[t]
\begin{center} \scalebox{1.2}[1.2]{\includegraphics{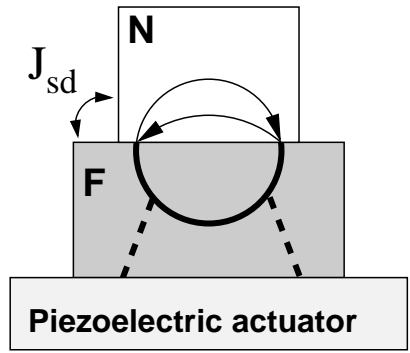}} \end{center} 
\caption{Schematic of the device structure used to detect the acoustic spin pumping~\cite{Uchida11,Uchida12a}. The dashed line represents the external phonon. The thin solid lines with arrows (bold lines without arrows) represent electron propagators (magnon propagators). 
} 
\label{fig_acousticSP01}
\end{figure}

\subsection{Acoustic spin pumping} 
To understand the role of nonequilibrium phonons in the spin Seebeck effect, it is instructive to discuss so-called acoustic spin pumping~\cite{Uchida11,Uchida12a} because the spin Seebeck effect is a kind of thermal spin pumping. In the acoustic spin pumping experiment, a hybrid structure of a ferromagnet $F$ and a nonmagnetic metal $N$ are attached to a piezoelectric actuator that acts as a nonequilibrium-phonon generator (see Fig.~\ref{fig_acousticSP01}). When nonequilibrium phonons are generated from the piezoelectric actuator and interact with magnons in the ferromagnet, the magnons deviate from the equilibrium distribution through magnon-phonon interaction and inject spin current into the nonmagnetic metal. We consider the interaction of exchange origin between magnons and phonons (the so-called volume magnetostrictive coupling~\cite{Bennett}), since this has been shown to give the largest contribution~\cite{Luthi}. The so-called single-ion magnetostriction~\cite{Bennett} arising from the spin-orbit interaction~\cite{Kittel} is assumed to be negligible, because if the latter coupling was relevant to the experiment in Refs.~\cite{Uchida11} and ~\cite{Uchida12a}, the resultant acoustic spin pumping should be seen at GHz frequencies instead of the MHz frequency at which the acoustic spin pumping is experimentally observed. However, we note that in the experiment of Ref.~\cite{Weiler12b}, the single-ion magnetostriction~\cite{Bennett} arising from the spin-orbit interaction~\cite{Kittel} seems to be dominant.

We start from the exchange Hamiltonian 
\begin{eqnarray}
  {\cal H}_{\rm ex} &=& -\sum_{\bmR_i, \bmR_j} J_{\rm ex}(\bmR_i - \bmR_j) \; 
  {\bm S}(\bmR_i) \cdot {\bm S}(\bmR_j) - \gamma \hbar {\bm H}_0 \cdot {\bm S}(\bmR_i), 
  \label{Eq:H_ex01}
\end{eqnarray}
where $J_{\rm ex}(\bmR_i -\bmR_j)$ is the strength of the exchange coupling between the ions at $\bmR_i$ and $\bmR_j$. The instantaneous position of the ion is written as $\bmR_i = \bmr_i + {\bm u}(\bmr_i)$ where the lattice displacement ${\bm u}(\bmr_i)$ is separated from the equilibrium position $\bmr_i$. Up to the linear order in the displacement, the exchange Hamiltonian (\ref{Eq:H_ex01}) can be written in the form 
\begin{equation}
  {\cal H}_{\rm ex} = \sum_\bmq \omega_\bmq a^\dag_\bmq a_\bmq + {\cal H}_{\rm mag-ph}, 
  \label{Eq:H_mag-ph00}
\end{equation} 
where 
$\omega_\bmq= \gamma H_0 + 2S_0 \sum_{\bm \delta} J_{\rm ex}({\bm \delta}) 
\sum_\bmq \big[ 1- \cos(\bmq \cdot {\bm \delta}) \big]$ 
is the magnon frequency with the lattice vector ${\bm \delta}=a_S \widehat{\bm \delta}$, and 
\begin{equation}
{\cal H}_{\rm mag-ph} = \sum_{\bmr_i, {\bm \delta}} 
({g \widehat{\bm \delta} }/{a_{\rm S}} )  \cdot  
\big[{\bm u}( \bmr_i)-{\bm u}(\bmr_i+ {\bm \delta}) \big] 
\bmS(\bmr_i) \bmS(\bmr_i+ {\bm \delta}) 
  \label{Eq:H_mag-ph01}
\end{equation}
is the magnon-phonon interaction with the magnon-phonon coupling $g$ given by $\nabla J_{\rm ex} ({\bm \delta}) = (g/a_S) \widehat{\bm \delta}$. 

In our case of acoustic spin pumping, the phonon is colored by a single wavenumber and frequency. The lattice displacement field ${\bm u}$ for a fixed wavenumber $\bmK_0$ is expressed as~\cite{Mahan} 
${\bm u} (\bmr_i,t) = \ui \sum_{\bmK=\pm \bmK_0} \widehat{\bm e}_{\bmK} 
U_\bmK(t) e^{\ui \bmK_0 \cdot \bmr_i} $, 
where the polarization vector 
$\widehat{\bm e}_{\bmK}$ is odd under the inversion $\bmK \to - \bmK$, and $U_\bmK(t)$ can be expressed as $U_\bmK(t)= u_{\bmK}(t)+ u_{-\bmK}(t)^*$ to satisfy $U_\bmK(t)=U_{-\bmK}(t)^*$. Note that the spatial average of $[u(\bmr_i)]^2 $ is given by $\langle [u(\bmr_i)]^2 \rangle_{\rm av} = 2 |U_{\bmK_0}|^2$. Using this representation of the displacement vector and introducing the magnon operator $a$, $a^\dag$ [Eqs.~(\ref{Eq:magnon-Sx01})-(\ref{Eq:magnon-Sz01})], the magnon-phonon interaction becomes 
\begin{equation}
    {\cal H}_{\rm mag-ph} = 
    \sum_{\bmq, \bmK=\pm \bmK_0} 
    \Lambda_{\bmK,\bmq} U_{\bmK} a^\dag_{\bmq+\bmK} a_\bmq , 
    \label{Eq:H_mag-sound02}
\end{equation}
where 
$\Lambda_{\bmK,\bmq}= \widetilde{g} \hbar \omega_\bmq 
(\bmK \cdot \widehat{\bm e}_{\bmK}) $ 
with 
$\widetilde{g}= 
\sum_{\bm \delta} 
\widehat{\bm \delta} \cdot {\bm \nabla}J_{\rm ex}({\bm \delta})
/[\sum_{\bm \delta} J_{\rm ex}({\bm \delta})]$ 
being the dimensionless magnon-phonon coupling constant. Note that, up to the lowest order in ${\bm u}$, only the longitudinal phonons couple to magnons when the phonons propagate along the symmetry axis of the crystal~\cite{Luthi}. 

Now we consider the process shown in Fig.~\ref{fig_acousticSP01}, in which nonequilibrium phonons interact with magnons and cause their nonequilibrium, thereby injecting a spin current into the attached nonmagnetic metal. As before, when we treat the phonon-dressed magnon lines as a single magnon propagator $\delta \check{X}_\bmq(\omega)$, it has the form 
\begin{eqnarray}
\delta \check{X}_\bmq(\omega) &=& 
\sum_{\bmK =\pm \bmK_0} 
\Lambda_{\bmK,\bmq}^2 |U_\bmK|^2 
\check{X}_\bmq (\omega) \check{X}_{\bmq - \bmK} (\omega - \nu_\bmK) 
\check{X}_\bmq (\omega), 
\label{Eq:Xnonloc02}
\end{eqnarray}
where $\nu_\bmK=v_p K$ is the phonon energy for the phonon velocity $v_p$. When we substitute Eq.~(\ref{Eq:Xnonloc02}) into Eq.~(\ref{Eq:I_s01}) and use Eq.~(\ref{Eq:C-func01}) with $\check{X}_\bmq(\omega)$ replaced by $\delta \check{X}_\bmq(\omega)$, we obtain the expression 
\begin{equation}
I_{\rm s} = \frac{\sqrt{2} \hbar (J_{\rm sd}^2 S_0) }{N_{\rm P} N_{\rm F}/N_{\rm int}}
  \sum_{\bmk,\bmq,\bmK=\pm \bmK_0} A_{\bmk,\bmq}(\nu_{\bmK}) 
  \Lambda_{\bmK,\bmq} |U_{\bmK}|^2 
  \label{Eq:spump-phonon01}
\end{equation}
for the acoustic spin pumping, where the quantity $A_{\bmk,\bmq}(\nu)$ is defined by 
\begin{eqnarray}
  A_{\bmk,\bmq}(\nu) &=& \int_\omega \; \Im {\rm m}\chi^R_\bmk (\omega) 
  \Im{\rm m}X^R_{\bmq-\bmK}(\omega-\nu) |X^R_\bmq(\omega)|^2 \nonumber \\
  && \qquad \times 
  \Big[\coth(\frac{\hbar (\omega-\nu) }{2 \kB T}) 
    - \coth(\frac{\hbar \omega}{2 \kB T}) \Big], 
  \label{Eq:def-A01}
\end{eqnarray}
which describes the correlation among the magnon, the phonon, and the itinerant spin density. Note that the acoustic spin pumping [Eq.~(\ref{Eq:spump-phonon01})] is proportional to the square of the phonon amplitude $|U_{\bmK}|^2$. Therefore, the acoustic spin pumping is proportional to the power of the external sound wave. 

\begin{figure}[t]
\begin{center}\scalebox{0.7}[0.7]{\includegraphics{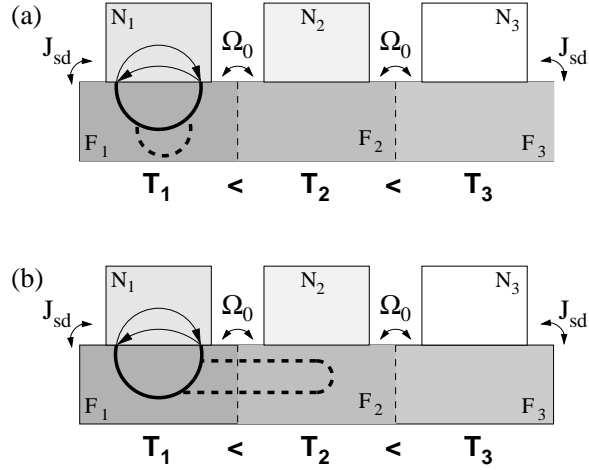}}\end{center} 
\caption{Feynman diagrams representing the phonon-drag contribution to the spin Seebeck effect~\cite{Adachi10}. The dashed line represents a phonon propagator. The thin solid lines with arrows (bold lines without arrows) represent electron propagators (magnon propagators). 
} 
\label{fig_ph-drag_diag01}
\end{figure}

\subsection{Phonon drag in the spin Seebeck effect}

In this subsection, we discuss the effect of nonequilibrium phonons on the spin Seebeck effect. In contrast to the previous subsection, the phonon in the this case is not an external field with a single color, but a statistical variable obeying Bose statistics. Therefore, it is necessary to represent the displacement field ${\bm u}$ with the phonon operator as 
\begin{equation}
{\bm u} (\bmr_i) = \ui \sum_{\bmK} 
{\widehat{\bm e}_{\bmK}}\sqrt{ \frac{\hbar}{{2 \nu_{\bmK} M_{\rm ion} N_{\rm F}}} }
\big( b_{\bmK} + b^\dag_{-\bmK} \big) e^{\ui \bmK \cdot \bmr_i}, 
\end{equation}
where $M_{\rm ion}$ is the ion mass and $b^\dag_{\bmK}$ ($b_{\bmK}$) is the phonon creation (annihilation) operator for wavevector $\bmK$, $\widehat{\bm e}_{\bmK}$ is the polarization vector, and $\nu_{\bmK}$ is the phonon frequency. Note that here and hereafter the polarization index $\zeta$ is omitted, because we consider a situation where $\zeta$ is not mixed with each other. Using this representation, the magnon-phonon interaction~(\ref{Eq:H_mag-ph01}) is expressed as 
\begin{equation} 
  {\cal H}_{\rm mag-ph} = 
  \frac{1}{\sqrt{N_{\rm F}}}   \sum_{\bmq,\bmK} 
  \Gamma_{{\bmK},\bmq} B_\bmK a^\dag_{\bmq+\bmK} a_\bmq, 
  \label{Eq:H_mag-ph02}
\end{equation}
where $B_\bmK= b_\bmK + b^\dag_{-\bmK}$ is the phonon field operator, and the magnon-phonon vertex is given by 
$\Gamma_{{\bmK},\bmq} = 2S_0 g \sum_{\bm \delta} 
\sqrt{\frac{\hbar \nu_\bmK}{2M_{\rm ion} v^2_{\rm p}}} 
(\widehat{\bm \delta}\cdot \widehat{\bm e}_\bmK ) 
(\widehat{\bm \delta} \cdot \widehat{\bmK} ) 
[ 1- \cos(\bmq \cdot {\bm \delta}) ]  $ 
with the phonon velocity $v_{\rm p}$. 

Now we discuss the phonon-drag contribution to the spin Seebeck effect. A natural guess is to replace $|U_{\bmK}|^2$ in Eq.~(\ref{Eq:spump-phonon01}) with the deviation of the phonon distribution function from its local equilibrium value, namely, $|U_{\bmK}|^2 \to <n_{\rm p}> - <n_{\rm p}>_{\rm loc-eq}$. In the following we show that this captures the essence of the phonon-drag contribution to the spin Seebeck effect. For illustration, let us first consider the process shown in Fig.~\ref{fig_ph-drag_diag01} (a), where the magnons emit and absorb phonons while traveling around the domain $F_1$, but neither the phonons nor magnons sense the temperature difference between $F_1$ and $F_2$. The phonon-dressed magnon propagator $\delta \check{X}_\bmq(\omega)$ in Fig.~\ref{fig_ph-drag_diag01} (a) can be expressed as 
\begin{eqnarray} 
  \delta \check{X}_\bmq(\omega) &=& 
  \check{X}_\bmq(\omega)  \check{\Sigma}_\bmq(\omega) \check{X}_\bmq(\omega) 
  \label{Eq:deltaX01} 
\end{eqnarray}
with the selfenergy due to phonons, 
\begin{eqnarray}
  \check{\Sigma}_\bmq (\omega) &=& 
  \frac{\ui}{2N_F} \sum_{\bmK} \left( \Gamma_{{\bmK},\bmq} \right)^2
  \int_\nu \Big\{ 
  D^R(\nu) \check{X}_{\bmq_-}(\omega_-) \check{\tau}_1 \nonumber \\
  && + D^A(\nu) \check{\tau}_1  \check{X}_{\bmq_-}(\omega_-) 
  + D^K(\nu) \check{X}_{\bmq_-}(\omega_-)   
  \Big\}, 
  \label{Eq:selfenergy01}
\end{eqnarray}
where $\check{{\bm \tau}}$ is the Pauli matrix in the Keldysh space, and we have introduced the shorthand notations $\omega_-= \omega-\nu$, $\bmq_-= \bmq- \bmK$, and $\int_\nu = \int_{-\infty}^\infty \frac{d \nu}{2 \pi}$. The bare phonon propagator in the above equation has the form 
\begin{equation}
  \widehat{D}_\bmK 
  =
  \left( { D^R_\bmK, \atop 0,} { D^K_\bmK \atop D^A_\bmK } \right), 
  \label{Eq:Dphonon01} 
\end{equation}
where the retarded component is given by 
$D^R_\bmK(\nu)= 
(\nu-\nu_\bmK + \ui /\tau_{\rm p})^{-1}- (\nu+\nu_\bmK+ \ui/\tau_{\rm p})^{-1}$ 
with $\tau_{\rm p}$ being the phonon lifetime~\cite{Mahan}, and each component satisfies the equilibrium condition: 
\begin{eqnarray}
D^A_\bmK(\nu)= [D^R_\bmK(\nu)]^*, \; \; 
D^K_\bmK(\nu)= 2 \ui \, {\rm Im} D^R_\bmK(\nu)
\coth(\frac{\hbar \nu}{2 \kB T} ). 
\label{Eq:D-eq01} 
\end{eqnarray} 

When the phonons are in thermal equilibrium, the phonon-dressed magnon propagator [Eq.~(\ref{Eq:deltaX01})] can be written in the local-equilibrium form 
\begin{eqnarray}
  \delta \check{X}_\bmq (\omega) &=& \delta \check{X}^{l \mathchar`-eq}_\bmq (\omega), 
  \label{Eq:Xloceq01}
\end{eqnarray}
where each component of the propagator satisfies the local-equilibrium condition, 
$\delta {X}_\bmq^{l \mathchar`-eq,A}(\omega) 
= [\delta {X}_\bmq^{l \mathchar`-eq,R}(\omega)]^*$ 
and 
$\delta {X}_\bmq^{l \mathchar`-eq,K}(\omega) = 
[\delta {X}_\bmq^{l \mathchar`-eq,R}(\omega)
-\delta {X}_\bmq^{l \mathchar`-eq,A}(\omega)]
\coth(\frac{\hbar \omega}{2 \kB T})$, 
with the retarded component given by 
\begin{eqnarray}
  \delta {X}_\bmq^{l \mathchar`-eq,R}(\omega) &=& 
   \sum_{\bmK} \frac{\ui (\Gamma_{{\bmK},\bmq})^2}{2N_F}
  \int_\nu  [X^R_\bmq(\omega)]^2 \Big\{ 
  D^R_\bmK(\nu)  \nonumber \\ 
  &\times& X^K_{\bmq_-}(\omega_-) 
  + D^K_\bmK(\nu) X^R_{\bmq_-}(\omega_-)  \Big\}. 
\end{eqnarray}
Using the same procedure to obtain Eq.~(\ref{Eq:Is_local01}), we calculate the injected spin current to be 
\begin{eqnarray}
  I_s &=& 
  -\frac{4 N_{\rm int} {J}_{\rm sd}^2 S_0}{\sqrt{2}\hbar^2 N_{N} N_{F}} 
  \sum_{\bmq,\bmk} 
  \int_{\omega} 
       \Im {\rm m} \chi_{\bmk}^{R}(\omega) 
       \Im {\rm m} \delta X_{\bmq}^{R}(\omega)  \nonumber \\
    && \quad \times 
       \left[ \coth(\frac{\hbar \omega}{2 \kB T_{N_1}}) 
         - \coth(\frac{\hbar \omega}{ 2 \kB T_{{F_1}}})   \right]. 
      \label{Eq:Is-phonon_local01} 
\end{eqnarray}
From this expression, we see that no spin current is injected into the nonmagnetic metal $N_1$ when $N_1$ and $F_1$ have the same temperature. 

The above result that the injected spin current vanishes when $T_{F_1}=T_{N_1}$ originates from the local-equilibrium condition of the magnons [Eq.~(\ref{Eq:Xloceq01})] which is derived from the equilibrium condition of the phonons [Eq.~(\ref{Eq:D-eq01})]. When the phonons deviate from thermal equilibrium, the corresponding phonon propagator $\delta \widehat{D}_\bmK(\nu)$ can be written in the form~\cite{Michaeli09} 
\begin{eqnarray} 
  \delta \widehat{D}_\bmK (\nu) &=& 
  \delta \widehat{D}^{l \mathchar`-eq}_\bmK (\nu) 
  + \delta \widehat{D}^{n \mathchar`-eq}_\bmK (\nu), 
  \label{Eq:Dnonloc01}  
\end{eqnarray}
where $\delta \widehat{D}_\bmK^{l \mathchar`-eq}(\nu)$ is the local-equilibrium propagator with local-equilibrium conditions 
$\delta {D}_\bmK^{l \mathchar`-eq,A}(\nu) 
= [\delta {D}_\bmK^{l \mathchar`-eq,R}(\nu)]^*$ and 
$\delta {D}_\bmK^{l \mathchar`-eq,K}(\nu) = 
[\delta {D}_\bmK^{l \mathchar`-eq,R}(\nu)-\delta {D}_\bmK^{l \mathchar`-eq,A}(\nu)] 
\coth(\frac{\hbar \nu}{2 \kB T})$, 
while $\delta \widehat{D}_\bmK^{n \mathchar`-eq}(\nu)$ describes the deviation from local equilibrium. When we allow such a nonequilibrium distribution of the phonons, the phonon-dressed magnon propagator cannot be expressed in the equilibrium form [Eq.~(\ref{Eq:Xloceq01})]. Instead, it is expressed in the form of Eq.~(\ref{Eq:Xnonloc01}) with the nonequilibrium component 
\begin{eqnarray} 
  \delta {X}^{n \mathchar`-eq,K}_\bmq (\omega) 
  &=&
  \sum_{\bmK} \frac{\ui (\Gamma_{\bmK,\bmq})^2}{2N_F} 
  \int_\nu    
  [X^R_\bmq(\omega_-) - X^A_\bmq(\omega_-)] |X^R_\bmq(\omega)|^2 \nonumber \\
  &\times& \delta D^{n \mathchar`-eq,K}_\bmK(\nu) 
  \left[ \coth(\frac{\hbar \omega_-}{2 \kB T_{F_1}}) ]
  - \coth(\frac{\hbar \omega}{2 \kB T_{{F_1}}})   \right]. 
    \label{Eq:dX_noneq02}
\end{eqnarray}
This nonlocal propagator can give rise to a nontrivial contribution to the injected spin current. 

With the above in mind, let us next consider the phonon-drag process shown in Fig.~\ref{fig_ph-drag_diag01} (b), where the phonons sense the temperature difference between $F_1$ and $F_2$ while the magnons do not. The phonon interaction between $F_1$ and $F_2$ is described by~\cite{Doniach74} 
\begin{eqnarray}
  {\cal H}_{\rm p}^{F \mathchar`- F} &=& 
  -\frac{1}{N_F} \sum_{\bmK,\bmK'} {\Omega}_{\rm p}^{\bmK + \bmK'} 
  B_\bmK \cdot B_{\bmK'}, 
\end{eqnarray}
where ${\Omega}_{\rm p}^{\bmK+\bmK'}$ is the Fourier transform of 
${\Omega}_{\rm p}(\bmr)= \Omega_0 
\sum_{\bmr_0 \in F \mathchar`- F \; {\rm interface}} 
a_S^3 \delta(\bmr-\bmr_0)$, 
and 
$\Omega_0 = \sqrt{2K_{\rm p}/M_{\rm ion}}$ with the elastic constant $K_{\rm p}$. The corresponding nonequilibrium phonon propagator $\delta \widehat{D}_\bmK(\nu)$ is given by 
\begin{eqnarray} 
  \delta \widehat{D}_\bmK (\nu)= 
  \frac{1}{N_F^2} \sum_{\bmK'} |\Omega_{\rm p}^{\bmK+\bmK'}|^2 
  \widehat{D}_\bmK(\nu)  \widehat{D}_{\bmK'}(\nu)  \widehat{D}_\bmK(\nu), 
\end{eqnarray}
which can then be written in the form of Eq.~(\ref{Eq:Dnonloc01}): 
$\delta \widehat{D}_\bmK^{l \mathchar`-eq} = 
({{\delta \widehat{D}_\bmK^{l \mathchar`-eq,R} }, \atop 
0,} 
{{2 \ui \Im {\rm m} [\delta \widehat{D}_\bmK^{l \mathchar`-eq,R}]} \atop 
[\delta \widehat{D}_\bmK^{l \mathchar`-eq,R}]^* })$ 
with 
\begin{eqnarray}
  \delta {D}_\bmK^{l \mathchar`-eq,R} (\nu) 
  &=& 
  \sum_{\bmK'}  \frac{|\Omega_{\rm p}^{\bmK+\bmK'}|^2}{N_F^2} 
    \big[{D}^R_\bmK(\nu) \big]^2  {D}^R_{\bmK'}(\nu) , 
\end{eqnarray}
and 
$\delta \widehat{D}_\bmK^{n \mathchar`-eq} = 
({0, \atop 0,} 
{ {\delta \widehat{D}_\bmK^{n \mathchar`-eq,K} } \atop 0})$ 
with 
\begin{eqnarray}
  \delta {D}_\bmK^{n \mathchar`-eq,K} (\nu) &=& 
  \sum_{\bmK'}  \frac{|\Omega_{\rm p}^{\bmK+\bmK'}|^2}{N_F^2} 
  [{D}^R_{\bmK'}(\nu)-{D}^A_{\bmK'}(\nu)] \nonumber \\
  &\times &
  |{D}^R_\bmK(\nu)|^2  
  \big[ \coth(\frac{\hbar \nu}{2 \kB T_{F_2}}) 
  - \coth(\frac{\hbar \nu}{2 \kB T_{{F_1}}}) \big] . 
  \label{Eq:dD_noneq01}
\end{eqnarray}

When we substitute Eq.~(\ref{Eq:dD_noneq01}) into Eq.~(\ref{Eq:dX_noneq02}) and use Eqs.~(\ref{Eq:C-func01}) and (\ref{Eq:I_s01}), we obtain the phonon-drag contribution to the injected spin current as 
\begin{eqnarray}
  I_{s}^{\rm drag} &=& 
  -  \frac{L}{N_N N_F^3} \sum_{\bmk,\bmq,\bmK,\bmK'}
  (\Gamma_{{\bmK},\bmq})^2 
  \int_\nu A_{\bmk,\bmq}(\nu) |{D}^R_\bmK(\nu)|^2 \nonumber \\
      & & \hspace{1cm} \times 
      {\rm Im}{D}^R_{\bmK'}(\nu) 
      \big[ \coth(\frac{\hbar \nu}{2 \kB T_{F_2}}) 
        - \coth(\frac{\hbar \nu}{2 \kB T_{{F_1}}}) \big] ,    
      \label{Eq:I_s02}
\end{eqnarray}
where $L= \sqrt{2} (J^2_{sd}S_0) \Omega_0^2 N_{\rm int} N'_{\rm int}/N_F$, and $A_{\bmk,\bmq}(\nu)$ is defined in Eq.~(\ref{Eq:def-A01}). After integrating over $\omega$ by picking up the magnon poles, $A_{\bmk,\bmq}(\nu)$ is calculated to be 
$A_{\bmk,\bmq}(\nu)= A_{\bmk,\bmq}^{(1)}(\nu)+ A_{\bmk,\bmq}^{(2)}(\nu)$ 
with 
$A_{\bmk,\bmq}^{(1)}(\nu)  = 
- \frac{1}{2}(e^{{\hbar \omega_\bmq}/{2 \kB T_{F_1}}}-1)^{-1}
\left( \frac{1}{\omega_\bmq} \chi_\bmk (\omega_\bmq) \right) 
\frac{\nu}{\nu^2 + 4\alpha^2 \omega_\bmq^2}$ 
and 
$A_{\bmk,\bmq}^{(2)}(\nu) = - \frac{1}{2}(e^{{\hbar \omega_\bmq}/{2 \kB T_{F_1}}}-1)^{-1}
\left( \frac{1}{\omega_\bmq} \chi_\bmk (\omega_\bmq) \right)^2 
(\frac{\omega_\bmq \tau_{\rm sf}}{\chi_N})  
\frac{ \nu^2} {\nu^2 + 4\alpha^2 \omega_\bmq^2}$. 
Note that owing to the symmetry in the $\nu$-integration, the leading term $A_{\bmk,\bmq}^{(1)}(\nu)$ does not contribute to the thermal spin injection. Then, we can perform the integration over $\nu$ by picking up the phonon poles, 
$\int_\nu |{D}^R_\bmK(\nu)|^2 
  {\rm Im}{D}^R_{\bmK'}(\nu) 
  \big[ \coth(\frac{\hbar \nu}{ \kB T_{F_2}}) 
        - \coth(\frac{ \hbar \nu}{\kB T_{{F_1}}}) \big] 
      = -{\pi \tau_{\rm p}} \delta(\nu_\bmK - \nu_{\bmK'}) 
      \big[ \coth(\frac{\hbar \nu_\bmK}{2 \kB T_{F_2}}) 
        - \coth(\frac{\hbar \nu_\bmK}{2 \kB T_{{F_1}}}) \big]$, 
which yields 
\begin{eqnarray}
  I_{s}^{\rm drag} &=& 
  \left( \frac{L \tau_{\rm p}}{4 \pi^3 \nu_{\rm D}^6} \right) \frac{1}{N_N N_F} 
  \sum_{\bmk,\bmq} 
  \int d \nu_\bmK \nu^4_\bmK \big(\Gamma_{\bmK,\bmq}\big)^2 \nonumber \\
  &\times& 
  A_{\bmk,\bmq}(\nu_\bmK) 
  \big[ \coth(\frac{\hbar \nu_\bmK}{2 \kB T_{F_2}}) 
    - \coth(\frac{\hbar \nu_\bmK}{2 \kB T_{{F_1}}}) \big], 
  \label{Eq:ph-drag_main01}
\end{eqnarray}
where $\nu_{\rm D}= v_{\rm p}/a_S$. 

The above expression, which is proportional to the phonon lifetime $\tau_{\rm p}$, gives the phonon-drag contribution to the spin Seebeck effect. After a rather lengthy calculation, Eq.~(\ref{Eq:ph-drag_main01}) is transformed into 
\begin{equation}
I_{\rm s} = k_{\rm B} (T_1-T_2) \left( \frac{ {\Gamma}_{\rm eff}^2 }{\hbar^2} \right) 
R {\cal B}\tau_{\rm p}, 
\label{Eq:Is_drag02} 
\end{equation}
where the dimensionless constant $\Gamma_{\rm eff}$ is given by 
${\Gamma}_{\rm eff}^2 = \left( \frac{\widetilde{g}^2 \hbar 
{\nu}_{\rm D}}{ M_{\rm ion} v_{\rm p}^2} \right) $, 
the factor 
$R= \frac{0.1 \times J_{\rm sd}^2 S_0 N_{\rm int} \chi_{\rm P}}
{\pi^2 (\lambda_{\rm sf}/a)^3(\Lambda/a_S)}$ 
measures the strength of the magnetic coupling at the F/N interface, and ${\cal B} = {\cal B}_1 \cdot {\cal B}_2$ where 
$ {\cal B}_1 = \frac{(T/T_{\rm D})^5}{4\pi^3}  \int_0^{T_{\rm D}/T} 
d u \frac{u^6}{\sinh^2 (u/2)}$ 
is a function of thermally-excited phonons with the Debye temperature $T_{\rm D}=\hbar \nu_{\rm D}/k_{\rm B}$, and 
$ {\cal B}_2  = \frac{(T/T_{\rm M})^{9/2} } {4 \pi^2} (\frac{\kB T_{\rm M} \tau_{\rm sf}}{\hbar})^3 \int_0^{T_{\rm M}/T} d v \frac{v^{7/2}}{e^{u}-1}$ 
is a function of thermally-excited magnons with $T_{\rm M}$ being the characteristic temperature corresponding to the magnon high-energy cutoff. 

The important point of Eq.~(\ref{Eq:Is_drag02}) is that the spin Seebeck signal due to phonon drag is proportional to the phonon lifetime $\tau_{\rm p}$, because the carriers of the heat current in this process are phonons. Because the phonon lifetime is strongly enhanced at low temperatures (typically below $100$ K) owing to a rapid suppression of the umklapp scattering, Eq.~(\ref{Eq:Is_drag02}) suggests that the spin Seebeck effect is enormously enhanced at low temperatures. In contrast, the signal at zero temperature should vanish because of the third law of thermodynamics. Therefore, the phonon-drag spin Seebeck effect must have a pronounced peak at low temperatures. Note that although the possibility of similar enhancement of the magnon lifetime in the magnon-driven spin Seebeck effect [Eq.~(\ref{Eq:Is_nonlocal03})] is not definitely excluded, judging from the ferromagnetic resonance linewidth in Y$_3$Fe$_5$O$_{12}$~\cite{Vittoria85} as a measure of the inverse magnon lifetime, it does not seem likely. 

To date, there are two experimental findings that support the existence of the phonon-drag spin Seebeck effect. The first is the observation of the predicted low-temperature peak in the temperature dependence of the spin Seebeck effect~\cite{Jaworski11,Uchida11b}. In Ref.~\cite{Adachi10} the earliest experimental data on the spin Seebeck effect in LaY$_2$Fe$_5$O$_{12}$ were theoretically analyzed, and the theory predicted that the spin Seebeck effect must show a pronounced peak at low temperatures as is discussed above. In Ref.~\cite{Jaworski11} the temperature dependence of the spin Seebeck effect was measured in (Ga,Mn)As, and the data showed a pronounced peak at low temperatures consistent with the theoretical prediction~\cite{Adachi10}. In Ref.~\cite{Uchida11b} the same trend was confirmed for YIG. The other experimental finding that supports the phonon-drag spin Seebeck effect is the observation of a spin Seebeck effect that is unaccompanied by a global spin current. Reference~\cite{Jaworski10} reported that cutting the magnetic coupling in (Ga,Mn)As while maintaining the thermal contact allowed the spin Seebeck effect to be observed even in the absence of global spin current flowing through (Ga,Mn)As. The phonon-drag spin Seebeck effect can explain the ``scratch'' test experiment~\cite{Jaworski10}, although the idea of a magnon-driven spin Seebeck fails to explain the experiment. Moreover, in a recent study~\cite{Uchida11}, an {\it isolated} NiFe alloy on top of a sapphire substrate was used to measure the spin Seebeck effect. This study excluded the possibility of a dipole-magnon-driven spin Seebeck effect for the ``scratch'' test experiment~\cite{Jaworski10}, and found that only the phonon drag by the substrate phonons could explain the experiment. One important point is that the experiment of Ref.~\cite{Uchida11} was performed at room temperature; nevertheless, the spin Seebeck effect was observed with the signal extended over several millimeters, as in the first observation of the spin Seebeck effect in NiFe alloy~\cite{Uchida08}. 
This result may suggest that the phonon drag can contribute to the spin Seebeck effect even at room temperature.

\section{Varieties of the spin Seebeck effect} 

\subsection{Longitudinal spin Seebeck effect}

\begin{figure}[t]
\begin{center}\scalebox{0.5}[0.5]{\includegraphics{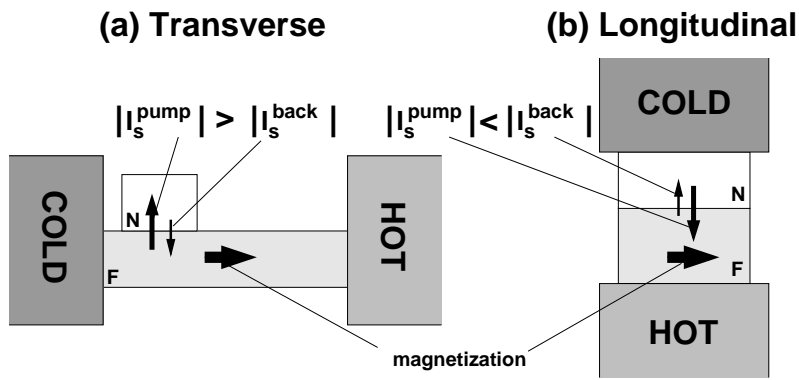}} \end{center} 
\caption{Schematic of the experimental setup for (a) the transverse spin Seebeck effect and (b) the longitudinal spin Seebeck effect.} 
\label{fig_LSSE-setup01}
\end{figure}

Up to this point, we have discussed the {\it transverse} spin Seebeck effect [Fig.~\ref{fig_LSSE-setup01} (a)], in which the direction of the thermal spin injection into a nonmagnetic metal is {\it perpendicular} to the temperature gradient. There is another type of spin Seebeck effect called the {\it longitudinal} spin Seebeck effect~\cite{Uchida10c,Uchida10b,Uchida11b} [Fig.~\ref{fig_LSSE-setup01} (b)], in which the direction of the thermal spin injection into a nonmagnetic metal is {\it parallel} to the temperature gradient. While both {\it conducting} and {\it insulating} ferromagnets can be used for the {transverse} spin Seebeck effect, the longitudinal spin Seebeck effect is well defined only for an {\it insulating} ferromagnet because of the parasitic contribution from the anomalous Nernst effect~\cite{Bosu11,Huang11,Weiler12a}. The longitudinal spin Seebeck effect has been observed in monocrystalline~\cite{Uchida10b} and polycrystalline~\cite{Uchida11b} YIG (Y$_3$Fe$_5$O$_{12}$) as well as in polycrystalline ferrite (Mn,Zn)Fe$_2$O$_4$~\cite{Uchida10c}. The longitudinal spin Seebeck effect is the simplest configuration in which a bulk polycrystalline ferromagnet can be used. Therefore, it is considered to be a prototype of the spin Seebeck effect from an application viewpoint. 

\begin{figure}[t]
\begin{center}
\includegraphics[width=10cm]{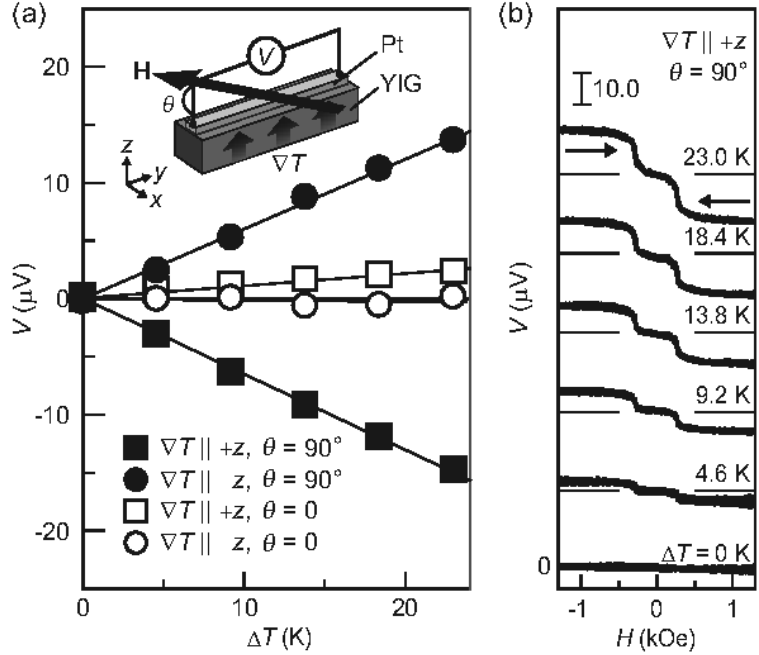} \end{center} 
\caption{(a) $\Delta T$ dependence of $V$ in the YIG/Pt sample at $H=1\,\textrm{kOe}$, measured when $\nabla T$ was applied along the $+z$ and $-z$ direction. The magnetic field ${\bf H}$ was applied along the $x$ direction ($\theta = 90^\circ$) and the $y$ direction ($\theta = 0$). (b) $H$ dependence of $V$ in the YIG/Pt sample for various values of $\Delta T$ at $\theta = 90^\circ$, measured when $\nabla T$ was along the $+z$ direction. }\label{fig:longitudinalSSE}
\end{figure}

In Fig. \ref{fig:longitudinalSSE}, we show typical experimental results for the longitudinal spin Seebeck effect. The sample consists of a monocrystalline YIG slab and a Pt film attached to a well-polished YIG (100) surface. 
The length, width, and thickness of the YIG slab are 6 mm, 2 mm, and 1 mm, while the corresponding dimensions of 
the Pt film are 6 mm, 0.5 mm, and 15 nm. An external magnetic field ${\bf H}$ (with the magnitude $H$) was applied in the $x$-$y$ plane at an angle $\theta$ to the $y$ direction (see Fig. \ref{fig:longitudinalSSE}(a)). A temperature difference $\Delta T$ was applied between the top and bottom surfaces of the YIG/Pt sample. Figure \ref{fig:longitudinalSSE}(a) shows the voltage $V$ between the ends of the Pt layer in the YIG/Pt sample as a function of $\Delta T$ at $H=1\,\textrm{kOe}$. When ${\bf H}$ was applied along the $x$ direction ($\theta = 90^\circ$), the magnitude of $V$ was observed to be proportional to $\Delta T$. The sign of the $V$ signal at finite values of $\Delta T$ is clearly reversed by reversing the $\nabla T$ direction. The $V$ signal also changes its sign with reversing ${\bf H}$ when $\theta = 90^\circ$ [Fig. \ref{fig:longitudinalSSE}(b)] and disappears when ${\bf H}$ is along the $y$ direction ($\theta = 0$) [Fig. \ref{fig:longitudinalSSE}(a)]. These results are consistent with the symmetry of the inverse spin Hall effect induced by the longitudinal spin Seebeck effect [see Eqs. (\ref{Eq:ISHE01}) and (\ref{Eq:ISHE02})]. 

A major feature of the longitudinal spin Seebeck effect is that the sign of the spin injection is opposite to that in the transverse spin Seebeck effect, as shown in Fig.~\ref{fig_LSSE-setup01}. 
Focusing on the spin current injected into the nonmagnetic metal ($N$) close to the cold reservoir, the magnitude of the pumping component $I_s^{\rm pump}$ is greater than that of the backflow component $I_s^{\rm back}$ in the transverse spin Seebeck effect. In contrast, the magnitude of $I_s^{\rm pump}$ is less than $I_s^{\rm back}$ in the longitudinal spin Seebeck effect. Note that magnons carry spin $-1$, such that the pumping and backflow components have a negative sign.

A linear-response approach to the longitudinal spin Seebeck effect was developed in Ref.~\cite{Adachi12}. Here, we present a phenomenological argument. First, recall that the spin Seebeck effect can be understood in terms of the imbalance between the thermal noise of the magnons in the ferromagnet and the thermal noise of the conduction-electron spin density in the nonmagnetic metal. The former noise injects the spin current into the nonmagnetic metal, while the latter ejects the spin current from the nonmagnetic metal. Because the thermal noise in each element can be related to its effective temperature through the fluctuation-dissipation theorem, the spin Seebeck effect can also be interpreted in terms of the imbalance between the effective temperature of the magnons in the ferromagnet and the effective temperature of the conduction-electron spins in the nonmagnetic metal [see Eq.~(\ref{Eq:Is04})]. 

Then, the signal sign reversal between the longitudinal and the conventional transverse spin Seebeck effects may be explained by the following conditions: (i) Most of the heat current in the ferromagnet/nonmagnetic-metal hybrid system at room temperature is carried by phonons (see discussion in Ref.~\cite{Slack71} in the case of YIG), and (ii) the interaction between the phonons and the conduction-electron spins in the nonmagnetic metal $N$ is much stronger than the magnon-phonon interaction in the ferromagnet $F$. In the longitudinal spin Seebeck experiment, the nonmagnetic metal is in direct contact with the heat bath, and thereby is exposed to the phonon heat current due to condition (i). Then, because of condition (ii), the conduction-electron spins in the nonmagnetic metal $N$ are heated up faster than the magnons in the ferromagnet $F$, and the effective temperature of the conduction-electron spins in the nonmagnetic metal rises above that of the magnons in the ferromagnet $F$. In the conventional spin Seebeck setup, by contrast, the nonmagnetic metal $N$ is out of contact with the heat bath, and the phonon heat current does not flow through the nonmagnetic metal $N$, while the ferromagnet $F$ is in contact with the heat bath. This results in an increase in the effective magnon temperature in the ferromagnet $F$. Therefore, in this case, the effective temperature of the conduction-electron spins in the nonmagnetic metal $N$ is lower than that of the magnons in the ferromagnet $F$. This difference can explain the sign reversal of the spin Seebeck effect signal between the longitudinal and transverse setups.

\subsection{Thermoelectric coating based on the spin Seebeck effect}

The spin Seebeck effect in magnetic insulators can be used directly to design thermo-spin generators and, combined with the inverse spin  Hall effect, thermoelectric generators, allowing new ways to improve thermoelectric generation efficiency. In general, the efficiency is improved by suppressing the energy losses due to heat conduction and Joule dissipation, which are realized respectively by reducing the thermal conductivity $\kappa$ for the sample part where heat currents flow and by reducing the electrical resistivity $\rho$ for the part where charge currents flow. In thermoelectric metals, the Wiedemann-Franz law ($\kappa_{\rm e} \rho=$~constant) limits this improvement in electric conductors when $\kappa$ is dominated by the electronic thermal conductivity $\kappa_{\rm e}$. A conventional way to overcome this limitation is to use semiconductor-based thermoelectric materials, where the thermal conductance is usually dominated by phonons while the electric conductance is determined by charge carriers and thus $\kappa$ and $\rho$ are separated according to the kind of the carriers. The spin Seebeck effect provides another way to overcome the Wiedemann-Franz law; in the spin Seebeck device, the heat and charge currents flow in different parts of the sample: $\kappa$ is the thermal conductivity of the magnetic insulator, and $\rho$ is the electrical resistivity of the metallic wire, such that $\kappa$ and $\rho$ in the spin Seebeck device are segregated according to the part of the device elements. Therefore, the spin Seebeck effect in insulators allows us to construct thermoelectric devices operated by an entirely new principle, although the thermoelectric conversion efficiency is still small at present.

\begin{figure}[t]
\begin{center}\includegraphics[width=11cm]{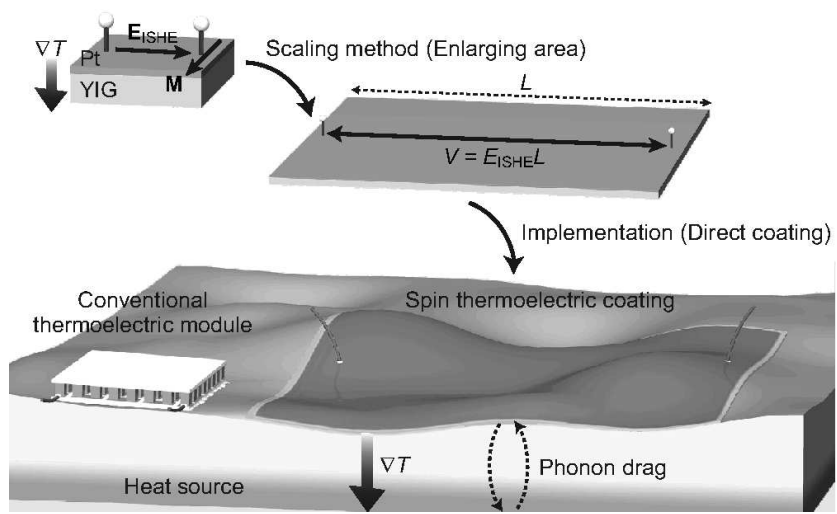}\end{center}
\caption{Concept of the STE coating based on the spin Seebeck effect~\cite{Kirihara12}. The STE coating exhibits a straightforward scaling: a larger film area leads to a larger thermoelectric output. Such a simple film structure can be directly coated onto heat sources with different shaped (curved or uneven) surfaces. }
\label{fig:kirihara}
\end{figure}

In 2012, Kirihara {\it et al.} proposed a new thermoelectric technology based on the spin Seebeck effect called ``spin-thermoelectric (STE) coating''~\cite{Kirihara12}, which is characterized by a simple film structure, convenient scaling capability, and easy fabrication (Fig. \ref{fig:kirihara}). In their experiments, an STE coating with a 60-nm-thick Bi-substituted YIG film was applied by using metal organic decomposition on a nonmagnetic substrate. Notably, thermoelectric conversion driven by the longitudinal spin Seebeck effect was successfully demonstrated under a temperature gradient perpendicular to such an ultrathin STE-coating layer (amounting to only 0.01 \% of the total sample thickness). The STE coating was found to be applicable even to glass surfaces with amorphous structures. Such a versatile implementation of thermoelectric function may give rise to other ways of making full use of omnipresent heat. 

\subsection{Position sensing via the spin Seebeck effect} 

\begin{figure}[t]
\begin{center}\includegraphics[width=6cm]{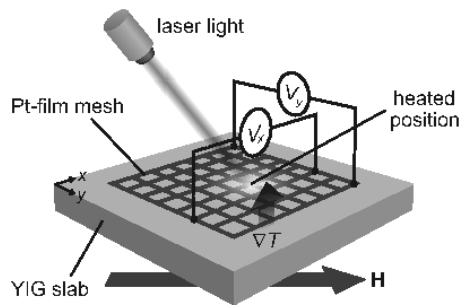}\end{center}
\caption{A schematic of the YIG-slab/Pt-mesh structure. In Ref.~\cite{Uchida11c}, part of the sample was heated by laser light, and the two-dimensional position information of the heated part was found by calculating the tensor product of the spatial profiles of the SSE voltage along the $x$ and $y$ directions. Here, an external magnetic field was applied along the diagonal ($45^\circ$) direction of the Pt mesh for generating the spin Seebeck voltage in both directions. } 
\label{fig:position} 
\end{figure}

The longitudinal spin Seebeck effect in magnetic insulators has also been used in two-dimensional position sensing using a YIG slab covered with a Pt-film mesh~\cite{Uchida11c}. Figure \ref{fig:position} shows a schematic of the YIG-slab/Pt-mesh sample. When part of the sample surface was heated, the position of the heated part was found from the spatial profile of the spin Seebeck voltage in the Pt mesh. The advantages of two-dimensional position sensing using the spin Seebeck effect are the simplicity of the device structure and the production cost; this device structure can be made simply by fabricating a patterned film on a commonly-used sintered polycrystalline insulator. Therefore, this position-sensing method gives us a realistic application of the spin Seebeck effect in thermally-driven user-interface devices and image-information sensors. 

\section{Other thermal spintronic effects} 

So far we have focused on the spin Seebeck effect. Besides the spin Seebeck effect, there are several intriguing phenomena in which the interplay of spin and heat plays a crucial role. In this section we briefly review other thermal spintronic effects. 

\subsection{Spin injection due to the spin-dependent Seebeck effect} 
A thermally driven pure spin-current injection across a charge-conducting interface has recently been reported by several groups, in which the ``spin-dependent Seebeck effect'' plays an important role. Slachter {\it et al.}~\cite{Slachter10} demonstrated thermally driven pure spin-current injection and its electrical detection using the nonlocal lateral geometry of NiFe/Cu. The physics behind this experiment is based on the spin-dependent thermoelectric effect. The spin-dependent current ${\bm j}_{\uparrow,\downarrow}$ is described by 
\begin{equation}
  {\bm j}_{\uparrow,\downarrow} = \sigma_{\uparrow,\downarrow} 
  \left( \frac{1}{e} {\bm \nabla} \mu_{\uparrow,\downarrow} + 
  S_{\uparrow,\downarrow} {\bm \nabla} T \right), 
\end{equation}
where $\sigma_{\uparrow,\downarrow}$, $\mu_{\uparrow,\downarrow}$, and $S_{\uparrow,\downarrow}$ are the spin-dependent conductivity, spin-dependent electrochemical potential, and spin-dependent Seebeck coefficient, respectively. The spatial distribution of the spin accumulation $\mu_\uparrow-\mu_\downarrow$ is described by the Valet-Fert spin diffusion equation: 
\begin{equation}
  \nabla^2 (\mu_\uparrow -\mu_\downarrow) 
  = \frac{1}{\lambda^2} (\mu_\uparrow -\mu_\downarrow), 
\end{equation}
where $\lambda$ is the spin-flip diffusion length. The essence of the experiment can be seen by solving these two equations under an appropriate temperature distribution across the NiFe/Cu interface. 

Le Breton {\it et al.}~\cite{Breton11} demonstrated thermal spin injection from NiFe into Si through an insulating tunnel barrier SiO$_2$/Al$_2$O$_3$ and called the phenomenon ``Seebeck spin tunnelling''. Here the injected spin current was detected by the Hanle effect, and the observed signal was analyzed in terms of the ``spin-dependent Seebeck effect''. It is important to note that the direction of the spin injection in these two experiments is parallel to the temperature gradient, such that the signal could contain the contribution from the longitudinal spin Seebeck effect. 

From the Kelvin relation $\Pi_{\uparrow,\downarrow}= TS_{\uparrow,\downarrow}$ with the spin-dependent Peltier coefficient $\Pi_{\uparrow,\downarrow}$, we expect the reciprocal process, i.e., the spin-dependent Peltier effect. Flipse {\it et al.}~\cite{Flipse12} have recently reported observation of this effect. 

\subsection{Seebeck effect in magnetic tunnel junctions} 
Several groups have measured the tunneling magneto-thermopower ratio of magnetic tunnel junctions, which was discussed analytically~\cite{Hatami09} and computed by a first-principles calculation~\cite{Czerner11}. Walter {\it et al.}~\cite{Walter11} and Liebing {\it et al.}~\cite{Liebing11} observed the tunneling magneto-thermopower in a CoFe/MgO/CoFe magnetic tunnel junction. The signal is caused by the spin-dependent Seebeck effect. 

\subsection{Magnon-drag thermopile} 
It is well known that two drag effects contribute to the thermoelectric effect in magnetic metals: one is the phonon drag in which nonequilibrium phonons transfer momentum to conduction electrons to produce thermopower, and the other is the magnon drag in which nonequilibrium magnons transfer momentum to conduction electrons~\cite{Blatt76}. However, the magnon-drag effect is easily masked by the phonon-drag effect, and in general, it is quite difficult to investigate only the magnon-drag effect. Costache {\it et al.}~\cite{Costache11} recently overcame this difficulty and proposed a device named the ``magnon-drag thermopile'' which provides information about the magnon-drag effect. The device is composed of many pairs of NiFe wires connected electrically in series with Ag wires, but placed thermally in parallel. When the two magnetizations in a pair of NiFe wires are in the parallel configuration, the thermopower is zero because the contributions of each wire are of the same magnitude but opposite signs. However, when the two magnetizations in a pair of NiFe wires are in the antiparallel configuration, there is a difference in the magnon states between the two wires, and the resultant thermopower is nonzero. Note that, in principle, although any electron-magnon scattering process other than the magnon drag can contribute to the observed thermopower, the magnon drag can dominate the signal when the energy dependence of the electron lifetime is negligible. 

\subsection{Thermal spin-transfer torque} 
Thermal spin-transfer torque is also a highly debated topic. Hatami {\it et al.}~\cite{Hatami07} discussed the thermal spin-transfer torque in magnetic nanostructures of metals, and Jia {\it et al.}~\cite{Jia11a} recently developed a first-principles estimation of the same process. This effect is relevant to the thermally driven domain wall motion discussed analytically by Kovalev {\it et al.}~\cite{Kovalev09} and computed numerically by Yuan {\it et al.}~\cite{Yuan10} Thermal spin-transfer torque has also been discussed in the context of magnetic insulators. Slonczewski~\cite{Slonczewski10} discussed the thermal spin-transfer torque resulting from the longitudinal spin Seebeck effect in ferrite. Spin-transfer torque caused by magnons is called magnonic spin-transfer torque~\cite{Yan11}, and Hinzke {\it et al.}~\cite{Hinzke11} discussed the role of thermal magnonic spin-transfer torque. Experimentally, an evidence for the thermal spin-transfer torque was reported by Yu {\it et al.}~\cite{Yu10}. 

\subsection{Effects of heat current on magnon dynamics} 

Another interesting subject is the dynamics of magnon wavepackets under the influence of a temperature gradient. Padr\'{o}n-Hern\'{a}ndez {\it et al.}~\cite{Rezende11} found that magnon wavepackets propagating along a YIG film are amplified when a temperature gradient is applied perpendicular to the YIG film. This experiment implies that the magnon damping term is canceled by the action of the temperature gradient, which leads to an amplification of the magnon wavepacket. The observed result was interpreted by the authors in terms of the magnonic spin-transfer torque of thermal origin in the longitudinal spin Seebeck configuration. 

Lu {\it et al.}~\cite{Lu12} studied the effects of heat current on ferromagnetic resonance. Using a trilayered structure consisting of a micron-thick YIG film grown on a submillimeter-thick gadolinium gallium garnet substrate and capped with a nanometer-thick platinum layer, they found that a temperature gradient over the trilayer can control the ferromagnetic relaxation in the YIG film. The result was interpreted by the authors in terms of the magnonic spin-transfer torque of thermal origin. 

\subsection{Anomalous Nernst effect and spin Nernst effect}
The anomalous Nernst effect refers to the generation of a voltage gradient ${\bm \nabla} V \parallel {\bm \hat{\bm m}} \times {\bm \nabla} T$ by applying a temperature gradient ${\bm \nabla} T$ in a ferromagnetic material with a magnetic polarization vector ${\bm \hat{\bm m}}$. This phenomenon has been studied systematically in various ferromagnetic metals by Miyasato {\it et al.}~\cite{Miyasato07}, in (Ga,Mn)As by Pu {\it et al.}~\cite{Pu08}, and in NiFe lateral spin valve by Slachter {\it et al.}~\cite{Slachter11}. It is important to note that if there is a thermal conductivity mismatch between the substrate and the ferromagnetic film when measuring the transverse spin Seebeck effect for a {\it conducting} magnet, there can be a parasitic contribution from the anomalous Nernst effect as pointed out in Ref.~\cite{Bosu11}. This issue was recently discussed again in Ref.~\cite{Huang11}. 

The spin Nernst effect refers to the generation of a transverse spin current $\bmJ_{\rm s}$ with the spin polarization $\widehat{\bm \sigma}$ by a temperature gradient, i.e., $\bmJ_{\rm s} \parallel \widehat{\bm \sigma} \times {\bm \nabla} T$. 
Reference~\cite{Cheng08} theoretically discusses the the spin Nernst effect in a two-dimensional Rashba spin-orbit system under a magnetic field, and Refs.~\cite{Liu10,Ma10} the same effect in a zero magnetic field. The spin Nernst effect of extrinsic origin is analyzed through first-principle calculations in Ref.~\cite{Tauber12}. 

\subsection{Thermal Hall effect of phonons and magnons} 
When the time-reversal symmetry is broken by a magnetic field or magnetic ordering, a finite Hall response can occur in principle even in the case of charge-neutral excitations such as phonons and magnons. Recently, the thermal Hall effect of phonons and magnons has been reported. Strohm {\it et al.} observed the thermal Hall effect of phonons in a paramagnetic insulator of terbium gallium garnet~\cite{Strohm05}. The result was explained by the interaction of local magnetic ions with the local orbital angular momentum of oscillating surrounding ions~\cite{Sheng06,Kagan08}. The thermal Hall effect of magnons is also observed in an insulating ferromagnet Lu$_2$V$_2$O$_7$ with pyrochlore structure~\cite{Onose10}, and the result was explained in terms of a Dzyaloshinskii-Moriya interaction. The Hall effect of magnons was also discussed theoretically in Refs.~\cite{Fujimoto09,Katsura10,Matsumoto11}. 

\section{Conclusions and future prospects} 
We have discussed the physics of the spin Seebeck effect and clarified the important role played by magnons. Moreover, we have shown that nonequilibrium phonons also play an active role. 
Below we summarize open theoretical and experimental questions in the spin Seebeck effect, as well as the directions of technical and industrial applications. 

One of the open theoretical questions in the spin Seebeck effect is the role of spin-polarized conduction electrons in metallic and semiconducting ferromagnets, especially in interpreting the experiment reported in Ref.~\cite{Jaworski12}. Another theoretical question is the existence of the reverse of the spin Seebeck effect, namely, the spin Peltier effect, which is different from the spin-dependent Peltier effect~\cite{Flipse12} and could be interpreted as a kind of magnonic Peltier effect from the viewpoint of the present article. In the magnon-driven spin Seebeck effect, a heat current in a ferromagnet drives the magnon spin current. On the other hand, if we rely on Onsager's argument on the symmetry of transport coefficients, we anticipate that the magnon spin current drives the heat current. The future challenges are to reveal the microscopic mechanism of the spin Peltier effect and to propose device structures for detecting this phenomenon. 

An open experimental question is a detection of the spin Seebeck effect at the compensation point of ferrimagnets that emerges from vanishing saturation magnetization, which was recently proposed~\cite{Ohnuma12}. In Ref.~\cite{Ohnuma12} the spin Seebeck effect in compensated ferrimagnets is theoretically investigated, and it is shown that the spin Seebeck effect survives even at the magnetization compensation point despite the absence of its saturation magnetization. This theoretical proposal awaits for experimental demonstrations. Another open experimental question is about the longitudinal spin Seebeck effect in a hybrid structure of a thin spin Hall electrode, {\it thin} magnet, and thick nonmagnetic substrate~\cite{Kirihara12,Rafa12}. In such a system, it is currently unclear whether a temperature gradient in the {\it thin} magnet is important or that in the thick substrate is important to the spin Seebeck effect, or a temperature difference across the magnet/spin-Hall-electrode interface is important. This issue is strongly related to practical applications and also related to the conventional thermoelectrics in superlattices~\cite{Mahan-book}, and hence should be investigated extensively. 

Regarding the direction of technical and industrial applications, the most important issue is to clarify to what extent the output power and efficiency can be enhanced. 
This requires at least three directions. The first is to construct a theoretical framework with which the maximum output power and efficiency can be discussed, as was done for conventional thermoelectrics~\cite{Ioffe-book}. The second is to maintain further material research to enhance the heat current/spin current conversion efficiency, giving a large spin current injection. The third is to develop a good spin-Hall electrode~\cite{Liu12,Niimi12} which can convert the injected spin current into a huge electric voltage. All of these efforts are necessary to achieve real industrial applications. Note that a small but a firm step is already in progress~\cite{Kirihara12,Uchida11c,Uchida12b}. 

Finally, one of the driving forces for investigating thermal effects in spintronics is the desire to deal with heating problems in modern solid-state devices. From this viewpoint, thermo-spintronics is still in its infancy, and many issues still remain unclear. For example, the relationship between the pure spin current and dissipation~\cite{Tulapurkar11} needs to be investigated extensively. Although the practical application of thermo-spintronics looks remote at present, we can definitely say that the interplay of spin and heat manifests itself in state-of-the-art experiments and involves interesting physics.

\ack 
We are grateful for fruitful discussions with S. Takahashi, J. Ohe, J. P. Heremans, G. E. W. Bauer, and T. An. 
This study was supported by a Grant-in-Aid for Scientific Research from the Ministry of Education, Culture, Sports, Science and Technology, Japan (MEXT), PRESTO-JST "Phase Interfaces for Highly Efficient Energy Utilization", CREST-JST "Creation of Nanosystems with Novel Functions through Process Integration", a Grant-in-Aid for Research Activity Start-up (24860003) from MEXT, a Grant-in-Aid for Scientific Research (A) (24244051) from MEXT, Japan, LC-IMR of Tohoku University, The Murata Science Foundation, The Mazda Foundation, and The Sumitomo Foundation.

\end{document}